\documentclass[journal]{IEEEtran}

\usepackage{verbatim}
\usepackage{optidef}
\usepackage{amssymb}
\usepackage[tableposition=top]{caption}
\usepackage{diagbox}

\let\llncssubparagraph\subparagraph
\let\subparagraph\paragraph
\usepackage[compact]{titlesec}
\let\subparagraph\llncssubparagraph

\usepackage{cite}
\ifCLASSINFOpdf
  \usepackage[pdftex]{graphicx}
  
  \DeclareGraphicsExtensions{.pdf,.jpeg,.png}
\else
 
  \usepackage[dvips]{graphicx}

 \DeclareGraphicsExtensions{.eps}
\fi
\usepackage{amsmath}
\usepackage{amssymb}
\usepackage{multirow}
\usepackage{array}
\usepackage{enumitem}
\newcolumntype{P}[1]{>{\centering\arraybackslash}p{#1}}
\newcolumntype{M}[1]{>{\centering\arraybackslash}m{#1}}
\titlespacing*{\section}
{0pt}{0pt}{0pt}
\titlespacing*{\subsection}
{0pt}{0pt}{0pt}
\titlespacing*{\subsubsection}
{0pt}{0pt}{0pt}

\begin{document}

\title{Risk-Informed Participation in T\&D Markets}

\author{Hafiz~Anwar~Ullah~Khan,~\IEEEmembership{Student Member,~IEEE}, Jip Kim,~\IEEEmembership{Student Member,~IEEE}, and Yury Dvorkin,~\IEEEmembership{Member,~IEEE}
      
}

\markboth{}%
{Shell \MakeLowercase{\textit{et al.}}: Bare Demo of IEEEtran.cls for IEEE Journals}

\maketitle

\begin{abstract}
Power producers can exhibit strategic behavior in electricity markets to maximize their profits. This behavior is more pronounced with the deregulation of distribution markets, which offers an opportunity for profit arbitrage between transmission and distribution (T\&D) markets. However, the temporally distinct nature of these two markets introduces a significant risk in profit for such producers. This paper derives its motivation from the perspective of a strategic producer and develops a Single Leader Multi-Follower (SLMF) game for deriving its participation strategies in T\&D markets, while accounting for different T\&D coordination schemes based on the individual market Gate Closure Times (GCT). We compare and contrast joint and sequential market clearing models with regulated and deregulated distribution environments and evaluate the risk of producer by leveraging consistent and coherent risk measures. SLMF game is reformulated as a Mathematical Program with Equilibrium Constraints (MPEC) and is solved using the seminal Scholtes's relaxation scheme. We validate the efficacy of our model and solution approach via the case study carried out on the 11-zone New York ISO, and 7-bus Manhattan power networks, used as transmission and distribution markets, respectively.
\end{abstract}

\IEEEpeerreviewmaketitle
\section*{Nomenclature}
\vspace{-5pt}
\subsection{Sets and Indices}
\begin{IEEEdescription}[\IEEEusemathlabelsep\IEEEsetlabelwidth{$\overline{\alpha}_{t,b},\underline{\alpha}_{t,b}$}]
\item[$b^{\textnormal{D}}_0$] {Root node of the distribution system}
\item[$B_{m/n} (b)$] {Set of ancestor/children buses of bus $b$, indexed by $b_{m/n}$ }
\item[$b^\textnormal{T}_c$]{Bus in the transmission network connected to $b^{\textnormal{D}}_0$ }
\item[\textbf{$B^{\textnormal{T/D}}$}] {Set of buses in the transmission/distribution network, indexed by $b^{\textnormal{T/D}}$}
\item[\textbf{$B^{\textnormal{T/D}}_{g}$}] {Set of buses containing generators in the transmission/distribution network, indexed by $b^{\textnormal{T/D}}$}
\item[$I^{\textnormal{T/D}}$]{Set of generators in transmission/distribution networks, indexed by $i^{\textnormal{T/D}}$}
\item[\textbf{$R$}] {Set of generators in the strategic producer, indexed by $r$}
\item[$T$] {Set of time intervals, indexed by $t$}
\item[$\wp$]{Set of random variables modeling DLMPs, indexed by $\wp_{b^\textnormal{D},t}$}
\end{IEEEdescription}

\subsection{Parameters}
\begin{IEEEdescription}[\IEEEusemathlabelsep\IEEEsetlabelwidth{$\overline{\alpha}_{t,b},\underline{\alpha}_{t,b}$}]
\item[$C_{i^{\textnormal{T/D}}}$] {Cost function of generator $i^{\textnormal{T/D}} \in I^{\textnormal{T/D}}$ }  
\item[$C_{r}$] {Cost function of the generator $r\in R$}
\item[$d_{b^\textnormal{T},t}$] {Total demand at bus $b^\textnormal{T}$, during time interval $t$}
\item[$d^\textnormal{{p/q}}_{b^\textnormal{D},t}$] {Total active/reactive power demand at bus $b^\textnormal{D}$, during time interval $t$}
\item[$F^{\textnormal{max}}_{(b^\textnormal{T},b_1)}$] {Maximum allowable power flow in transmission line connecting $b^\textnormal{T}$ and $b_1 \in B^\textnormal{T}$}
\item[$G^{{\textnormal{max/min}}}_{b^{\textnormal{T/D}}}$] {Maximum/minimum power output of generator connected to bus $b^{\textnormal{T/D}}$ at time interval $t$}
\item[$G_r^{\textnormal{max/min}}$] {Maximum/minimum power output of generator $r$ at time interval $t$}
\item[$Q^{{\textnormal{max/min}}}_{b^\textnormal{D}}$] {Maximum/minimum reactive power output of generator connected to bus $b^\textnormal{D}$ at time interval $t$}
\item[$S_{(b^\textnormal{D},b_1)}$] {Maximum apparent power flow in distribution feeder between $b^\textnormal{D}$ and $b_1 \in B^\textnormal{D}$}
\item[$U^{{\textnormal{max/min}}}_{b^\textnormal{D}}$] {Maximum/minimum square of voltage magnitude at bus $b^\textnormal{D}$ at time interval $t$}
\item[$X_{(b^\textnormal{T},b_1)}$] {Resistance of transmission line connecting $b^\textnormal{T}$ and $b_1 \in B^\textnormal{T}$}
\item[$x_{(b^\textnormal{T},b_1)}$] {Reactance of transmission line connecting $b^\textnormal{T}$ and $b_1 \in B^\textnormal{T}$}
\item[$\pi^\textnormal{D}_t$]{Time-of-use tariff in the distribution system at time interval $t$}
\end{IEEEdescription}

\subsection{Variables}
\begin{IEEEdescription}[\IEEEusemathlabelsep\IEEEsetlabelwidth
{$\overline{\alpha}_{t,b},\underline{\alpha}_{t,b}$}]
\item[$f_{(b^\textnormal{T},b_1),t}$] {Power flow in transmission line between $b^\textnormal{T}$ and $b_1 \in B^\textnormal{T}$ during time interval $t$}
\item[$f^{\textnormal{p/q}}_{(b^\textnormal{D},b_m),t}$] {Active/reactive power flow in distribution feeder between $b^\textnormal{D}$ and $b_m$ during time interval $t$}
\item[$g_{b^{\textnormal{T/D}},t}$] {Power output of generator connected to bus $b^{\textnormal{T/D}}$ at time interval $t$}
\item[$g_{i^{\textnormal{T/D}},t}$]{Power output of generator $i$ for transmission/distribution market for time interval $t$}
\item[$(g/q)_{b^\textnormal{D}_0,t}$] {Active/reactive power output of generator connected to the root node of distribution system, at time interval $t$}
\item[$g^{\textnormal{T/D}}_{r,t}$] {Power output of generator $r$ cleared by transmission/distribution market for time interval $t$}
\item[$g^{\textnormal{T}_o/\textnormal{D}_o}_{r,t}$] {Power output of generator $r$ offered by the producer to transmission/distribution network at time interval $t$}
\item[$q_{b^\textnormal{D},t}$] {Reactive power output of generator connected to bus $b^\textnormal{D}$, at time interval $t$}
\item[$u_{b^\textnormal{D},t}$] {Square of voltage magnitude at bus $b^\textnormal{D}$ at time interval $t$}
\item[$\lambda^{\textnormal{T/D}}_{b^{\textnormal{T/D}},t}$] {Market clearing price of wholesale/distribution market at bus $b^{\textnormal{T/D}}$ for time interval $t$}
\item[$\theta_{(b^\textnormal{T},b_{m/n}),t}$] {Voltage phase angle at bus $b^\textnormal{T}$ with respect to bus $b_{m/n}$ at time interval $t$}
\end{IEEEdescription}

\section{Introduction}
\label{intro}
\begin{figure}[!t]
\centering
\includegraphics[width=3.3in]{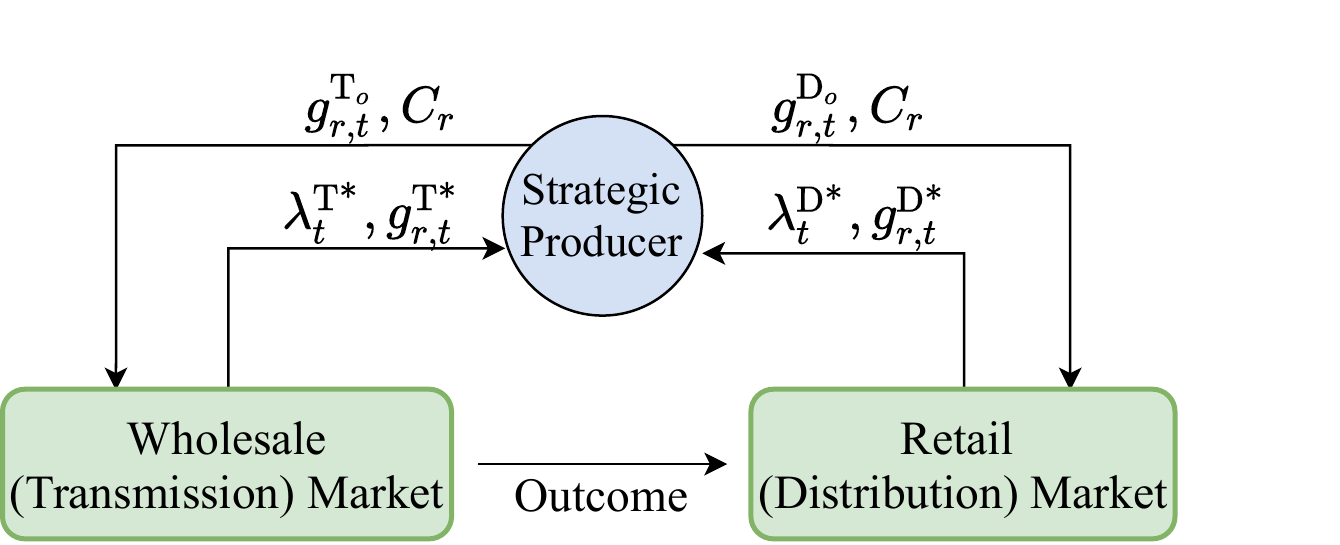}
\caption{\small{A schematic representation of the joint WM and DM.}}
\label{joint}
\vspace{-20pt}
\end{figure}

Liberalization of power systems and deployment of distributed energy resources at low and medium voltage levels provides a new impetus for rolling-out  distribution electricity markets. For instance, New York's Reforming the Energy Vision (NYREV) \cite{nyrev} is one of the most ambitious programs to overhaul electric power distribution that envisions a T\&D market structure, which will replace  the current  distribution environment predominantly  using regulated tariffs \cite{tariff} with a competitive market clearing mechanism for distribution networks, providing  producers with an opportunity to strategically sell electricity and ancillary services  in the two temporally and spatially distinct wholesale (transmission) and retail (distribution) marketplaces, as shown in Fig. \ref{joint}. However, this  shift from a regulated  environment to a competitive T\&D  marketplace will also impose profit risks  for  producers in distribution markets (DM) as their payoff will depend on uncertain market outcomes and will no longer be a priori known. Hence, this paper seeks to investigate risk-informed participation strategies of such strategic producers in emerging T\&D markets, while considering different  design options for distribution marketplaces and T\&D coordination.   

Traditionally, Stackelberg games have been adopted as primary modeling tools for  analyzing interactions between electricity producers and markets, \cite{dsm,stoch,gen_exp,storage}. Most of the existing literature, however, limits  hierarchical interactions among leaders (producers) and followers (markets) only to a single follower, owing to the current existence of only wholesale markets (WM), and the simplicity associated with manipulating its unique strategy \cite{SLMF_1}, \cite{Leyffer}. On the other hand, extensions with multiple followers that would allow for modeling both WM and DM are rare due to the complexity of solving Single-Leader-Multi-Follower (SLMF) games. Basilico \textit{et al.} \cite{SLMF_1} presented multiple mixed-integer nonlinear programming (MINLP) techniques for solving SLMF games, whereas a distributed prospect-theoretic solution algorithm is demonstrated in \cite{dsm}. These  MINLP techniques  reformulate  complementarity constraints using support vectors and treat resulting nonlinearities with convex envelopes. Similarly, \cite{SLMF_2} analyses  SLMF games for  optimistic and pessimistic cases that maximize and minimize the utility of the leader, respectively. The authors present an exact non-convex formulation for the optimistic case of normal-form and polymatrix games, and a heurisitic algorithm for solving the pessimistic case. Basilico \textit{et al.}  \cite{SLMF_2} motivate the application of heuristics for solving pessimistic SLMF games by the inapplicability of equivalent single-level reformulations. However, \cite{SLMF_3} casts the bilevel pessimistic normal-form SLMF game into a single-level mathematical program at the expense of replacing a supremum with a maximum, which sacrifices  optimality, and solves  this single-level program using a branch-and-bound algorithm.   

\begin{figure}[!t]
\centering
\includegraphics[width=3.5in]{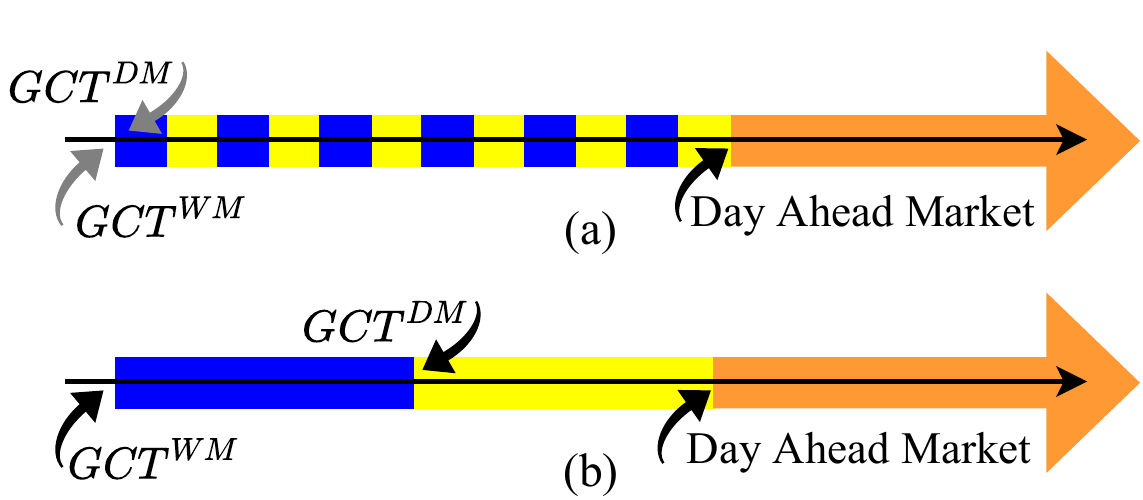}
\vspace{-15pt}
\caption{\small{Gate closure timeline for T\&D markets: (a) Simultaneous case, (b) Sequential case.}}
\label{time}
\vspace{-20pt}
\end{figure}

\begin{figure}[!t]
\centering
\includegraphics[width=3.3in]{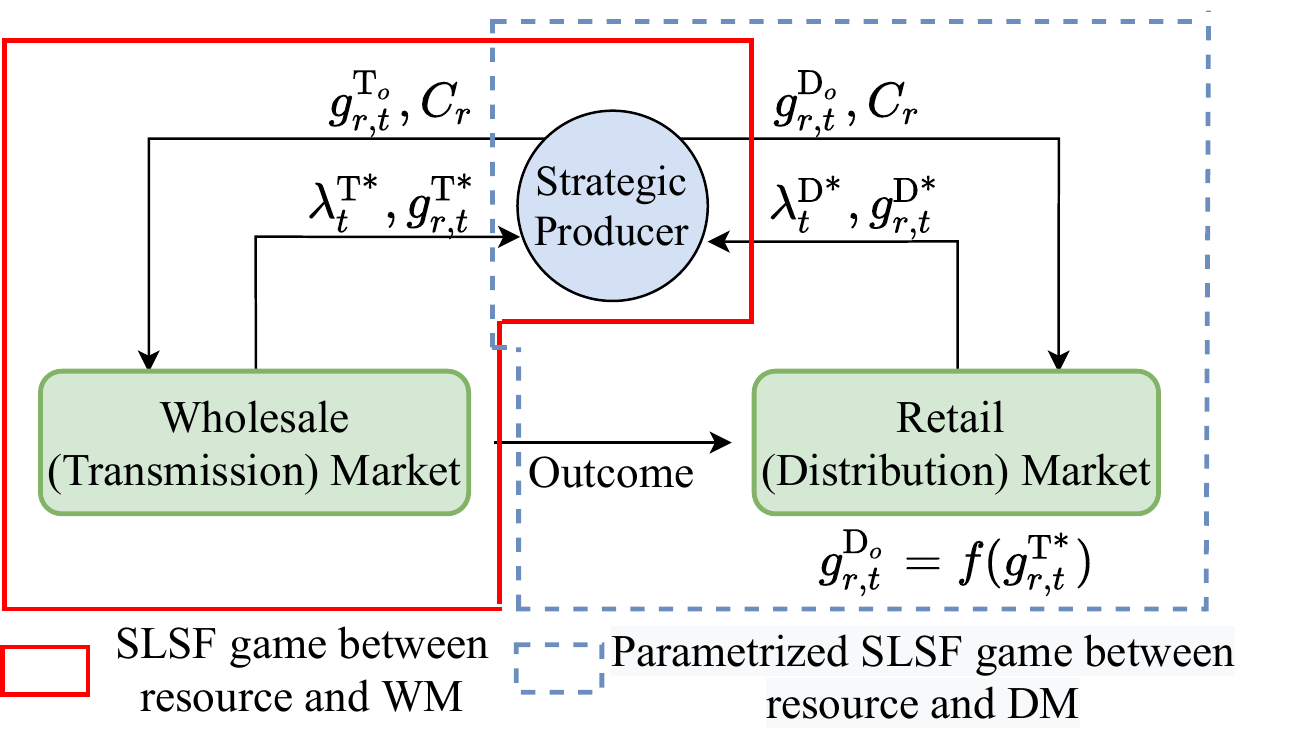}
\vspace{-10pt}
\caption{\small{Sequential formulation for SLMF games.}}
\label{seq}
\vspace{-20pt}
\end{figure}

Unlike established and somewhat generalizable WMs, DMs are still emerging and there is no consensus design, especially on the coordination between them, which in turn  depends on the sequence of their respective Gate Closure Times (GCT). Hence, we consider simultaneous and sequential GCTs of T\&D markets, see Fig. \ref{time}.  The simultaneous case is an idealization, which assumes GCT\textsuperscript{WM} and GCT\textsuperscript{DM} are concurrent, i.e. there is no opportunity for a producer to strategically arbitrage between profit opportunities in the WM and DM.  Alternatively, a more practical scenario is the sequential case which assumes that GCT\textsuperscript{WM} precedes GCT\textsuperscript{DM} \cite{caramanis}, thus rendering the DM strategy of the producer dependent on  its offer in WM. This dependency motivates the use of sequential SLMF games, shown in Fig. \ref{seq}. The first phase of this formulation is the SLSF game between a producer and the WM, whereas the second phase is the SLSF game between the producer and the DM, parametrized in the outcomes of the first phase. In both the simultaneous and sequential cases, we model a deregulated and a regulated environment for the distribution network.  The regulated case implies that there is an a priori time-of-use tariff \cite{tariff} that eliminates the risk of profit uncertainty for the producer. On the other hand, in the deregulated case, there is  uncertainty from the producer's perspective, since at GCT\textsuperscript{WM} the producer is unaware of the Distribution Locational Marginal Prices (DLMPs) and must account for this uncertainty while deriving its participation strategy, which  introduces profit risks for the producer. 

To internalize the effect of uncertain DLMPs, we exploit risk-informed optimization approaches that leverage historical data to reformulate the stochastic problem in a deterministic form. These approaches, including chance constraints (CC) and $\epsilon$-Conditional-Value-at-Risk ($\textnormal{CVaR}_\epsilon$), not only provide an accurate modeling mechanism for uncertainty, but also offer an insight to the associated risk tolerance \cite{cc}. The quantification of this uncertainty warrants the use of risk measures that evaluate the probability of uncertain outcomes (e.g. expected losses) to not exceed a pre-defined threshold, with respect to an underlying probability distribution. One such measure is $\epsilon$-Value-at-Risk ($\textnormal{VaR}_\epsilon$), which also implicitly allows enforcing CC \cite{robert}, as explained in Section \ref{reform}. However, $\textnormal{VaR}_\epsilon$ by definition, provides only a lower bound for losses in the tail of the distribution that renders this measure incapable of quantifying losses that might be incurred beyond the indicated threshold. This approach, therefore, elucidates an optimistic measure of risk, thus motivating a pursuit of more robust risk measures \cite{cvar_1}. For example, $\textnormal{CVaR}_\epsilon$ provides an estimate of the mean value of the $\epsilon$-tail distribution of random variables, constitutes a coherent risk measure, and maintains consistency with $\textnormal{VaR}_{\epsilon}$ in limited settings \cite{robert_cvar}. Hence, in this paper, we use CC and $\textnormal{CVaR}_{\epsilon}$ as  measures for profit risks  faced by the producer due to uncertain DLMPs.

One of the recurring strategies for solving Stackelberg games leverages KKT conditions or duality theory to reformulate the bilevel program into a single level equivalent formulation \cite{kktduality}. The KKT reformulation introduces complementarity conditions in the problem, giving rise to mathematical programs with equilibrium constraints (MPECs) for SLMF games. Hence, solution techniques developed for MPECs can be used to solve SLMF games if the lower-level optimization satisfies Slater's constraint qualification \cite{mpec_book}. MPECs are nonlinear programs (NLP) that (in addition to generally being NP-hard) do not satisfy the standard Mangasarian-Fromovitz constraint qualification and, therefore,  the stronger linear independence constraint qualification at any feasible point, invalidating convergence assumptions of standard NLP solution methods \cite{2_kanzow}. To overcome this limitation, multiple methods are available in literature for solving MPECs \cite{mpec_book} with different treatments of complementarity constraints. These methods include NLP reformulations including relaxation and penalization, combinatorial techniques employing branch-and-cut algorithms and active set methods, and implicit methods for solving MPECs \cite{mpec_book}. This paper  focuses on NLP reformulations and  relaxation techniques for MPECs to leverage off-the-shelf NLP solvers.

Since the seminal work of Scholtes \cite{scholtes} on relaxing the complementarity constraints of MPECs, various relaxation schemes \cite{lin, kadrani,ulbrich,jean}  have been introduced to transform an MPEC problem into a standard NLP form that can be solved using NLP solvers. As MPECs do not satisfy the standard constraint qualifications for NLPs, KKT conditions cannot be regarded as appropriate stationarity concepts; hence depending on the nature of bi-active sets, various weaker stationarity concepts, such as C, M, B, and strong stationary points are defined in this case \cite{relaxation_comp}. The convergence of relaxed NLPs to one of these points defines the proximity of the obtained solution to the actual optimizer \cite{relaxation_comp}. Relaxation approaches  \cite{lin, kadrani,ulbrich,jean} therefore  enhance  theoretical properties of relaxed NLPs, such as convergence to a stronger stationary point. However, numerical results demonstrate that the proposed relaxation techniques do not converge to a theoretically  stronger stationary point, rather converge to an inexact one, leading to a weaker stationarity result. The relaxation scheme by Scholtes, on the contrary, converges to a C-stationary point even if the convergence is obtained to an inexact stationary point \cite{inexact}. This characteristic is superior to other proposed regularization methodologies and motivates the application of this scheme for solving MPECs in this paper.

To recapitulate the above discussion, moving from a regulated distribution environment to a competitive DM exposes strategic producers to different levels of risk in their profit. This risk is a direct consequence of different sequences of GCTs emanating from the T\&D coordination in this emerging market structure. The strategic producer would therefore derive such risk-informed participation strategies for the WM and DM that maximize its profit. This paper presents a framework for the formulation and analysis of these participation strategies incorporating multiple GCT-based bi-level optimization formulations and an MPEC-relaxation based solution algorithm for the resulting SLMF game, accounting for different market environments at the distribution level. We employ risk-informed optimization to capture the uncertainties associated with DLMPs and co-optimize WM and DM decisions.


\section{Problem Formulation}
In this section, we provide a mathematical formulation for each player of the SLMF game. These formulations apply to the deregulated distribution environment, whereas for the regulated environment, we modify parts of these formulations as described in Remark 1.

\subsection{Formulation of the Producer}
We model the producer as a group of generators that maximizes its profit in the  DM and WM. This producer is  the leader in the SLMF game, and its decision-making process is formalized as follows:
\begin{subequations}
\label{leader}
\begin{IEEEeqnarray}{C}
\label{20}
\begin{split}
\text{max} \sum_{r \in R, t \in T}{(\lambda^\textnormal{T}_{b^\textnormal{T},t} } - C_r) g^{\textnormal{T}_o}_{r,t} + \hspace{-10pt} \sum_{r \in R, t \in T}{(\lambda^\textnormal{D}_{b^\textnormal{D},t}} - C_r) g^{\textnormal{D}_o}_{r,t}
\end{split}
\vspace{-5pt}
\end{IEEEeqnarray}
\vspace{-8pt}
subject to
\begin{IEEEeqnarray}{rCl}
\label{21}
G_r^{\textnormal{min}} & \leq & g^{\textnormal{T}_o}_{r,t} + g^{\textnormal{D}_o}_{r,t} \leq G_r^{\textnormal{max}}; \hspace{6mm} \forall r \in R, t \in T
\end{IEEEeqnarray}
\end{subequations}

Eq. (\ref{20}) maximizes the total profit of the strategic player in terms of power offered to the DM and WM. The objective is constrained by the  power production limits of the producer, as in eq. \eqref{21}. 

\subsection{Formulation of the Wholesale Market}
We assume that the wholesale market is operated over a meshed transmission network and use the  DC power flow assumptions as follows:
\begin{subequations}
\label{lmnopq}
\begin{IEEEeqnarray}{C}
\label{1}
    \text{min} \sum_{i^\textnormal{T} \in I^\textnormal{T},t \in T}{C_{i^\textnormal{T}}}g_{i^\textnormal{T},t} + \sum_{r \in R,t \in T}{C_{r} g^\textnormal{T}_{r,t}}
    \vspace{-10pt}
\end{IEEEeqnarray}
subject to \{
\begin{IEEEeqnarray}{ccc}
\label{5}
\begin{split}
    &\sum_{b^\textnormal{T} \in B^\textnormal{T}_{g}}(g_{b^\textnormal{T},t}) +\hspace{-10pt} \sum_{b_m \in B_m(b^\textnormal{T})}{f_{(b^\textnormal{T},b_m),t}} +\sum_{r \in R}{g^\textnormal{T}_{r,t}}=\hspace{-10pt}\sum_{b_n \in B_n(b^\textnormal{T})}{f_{(b^\textnormal{T},b_n),t}}\\[-3pt]&+ \sum_{b^\textnormal{T} | b^\textnormal{T} = b^\textnormal{T}_c}{f_{(b^\textnormal{T},b_o^\textnormal{D}),t}} + d_{b^\textnormal{T},t} : (\lambda^\textnormal{T}_{b^\textnormal{T},t}); \hspace{2 mm} \forall b^\textnormal{T} \in B^\textnormal{T}
\end{split}
\end{IEEEeqnarray}
\vspace{-15pt}
\begin{IEEEeqnarray}{c}
\label{2}
G_{i^\textnormal{T}}^{{\textnormal{min}}} \leq  g_{i^\textnormal{T},t} \leq G_{i^\textnormal{T}}^{{\textnormal{max}}} : (\underline{\gamma}_{i^\textnormal{T},t}, \overline{\gamma}_{i^\textnormal{T},t}) \hspace{2 pt}; \forall i^\textnormal{T} \in I^\textnormal{T}\\
\label{2.5}
0 \leq {g^\textnormal{T}_{r,t}} \leq g^{\textnormal{T}_o}_{r,t} :(\underline{\alpha}_{r,t}^\textnormal{T},\overline{\alpha}_{r,t}^\textnormal{T}); \hspace{2mm} \forall r \in R\\
\label{3}
\begin{split}
 f_{(b^\textnormal{T},b_1),t} = \frac {\theta_{(b^\textnormal{T},b_m),t} - \theta_{(b_1,b_n),t}} {x_{({b^\textnormal{T}},  {b_1})}} : (\xi_{(b^\textnormal{T},b_1),t}); \\[-2pt] \hspace{3 mm}\forall b^\textnormal{T}, b_1 \in B^\textnormal{T}, b_m \in B_m(b^\textnormal{T}), b_n \in B_n(b^\textnormal{T})
\end{split}
\end{IEEEeqnarray}
\vspace{-11pt}
\begin{IEEEeqnarray}{c}
\begin{split}
\label{4}
 -F_{(b^\textnormal{T},b_1)}^{\textnormal{max}} \leq  f_{(b^\textnormal{T},b_1),t} \leq  F_{(b^\textnormal{T},b_1)}^{\textnormal{max}} : \\\hspace{-3 pt}(\underline{\delta}_{(b^\textnormal{T},b_1),t}, \overline{\delta}_{(b^\textnormal{T},b_1),t}); \hspace{3 mm}\forall b^\textnormal{T}, b_1 \in B^\textnormal{T}\}, \forall t \in T
 \end{split}
\end{IEEEeqnarray}
\end{subequations}

Eq. \eqref{1} minimizes the total cost of power production from the generators in the transmission network, including the generators operated by strategic producer $r \in R$. Constraint \eqref{5} expresses the nodal power balance in the system (where $f_{(b^T,b_o^D),t} \coloneqq g_{b^D_{0},t}$  is the interface power flow between the transmission and distribution networks  defined in eq.~\eqref{distribution} of the  DM model), whereas constraint \eqref{2} limits the power of each generator with  its minimum and maximum generation limits. The power flow of each transmission line is calculated in  \eqref{3} and \eqref{4} enforces transmission  flow  limits. Dual variables  are provided in parenthesis after  each constraint.

\subsection{Formulation of the Distribution Market}
The objective function of the DM in eq. \eqref{6} is to minimize the overall generation cost  in the distribution network, including the production cost of distribution-level generation resources and  cost of the interface flow  between the WM and DM.  Electric power distribution can be modeled using a set of linearized AC power flow equations based on the \textit{LinDistFlow} formulation \cite{lindist}. These equations pertain to the extraction of all active and reactive power from the distribution substation given by constraints (\ref{7}) and  (\ref{8}), nodal active and reactive power balance at all buses expressed by constraints (\ref{9}) and (\ref{91}), capacity constraints for generating units in the distribution system and the strategic producer modeled by (\ref{10}), (\ref{14}), and (\ref{res}), and  voltage and power flow limits on each bus and feeder of the system expressed by constraints (\ref{12}), \eqref{12.1},  and (\ref{13}), respectively. Constraint \eqref{12.1} limits the interface power between the transmission and distribution systems, whereas \eqref{13} constrains the active and reactive power flow on feeders with their apparent power flow limit \cite{lindist}. Note that  \eqref{13} is a conic constraint, where $K := \{x \in \mathbb{R}^3 \vert x_1^2 \geq x_2^2 +x_3^2\}$ and  $K^\ast$  denote primal and dual second-order cones \cite{boyd}.  Finally, the nodal voltage magnitudes are modeled in eq.~\eqref{121}.  

\begin{subequations}
\label{distribution}
\begin{IEEEeqnarray}{C}
\label{6}
\begin{split}
\text{min} \sum_{i^\textnormal{D} \in I^\textnormal{D},t \in T}{C_{i^\textnormal{D}}g_{i^\textnormal{D},t}} +\hspace{-10 pt}\sum_{r \in R,t \in T}{C_{r} g^\textnormal{D}_{r,t}} + \hspace{-1 pt}\lambda^\textnormal{T}_{b^\textnormal{T},t} g_{b^\textnormal{D}_{0},t}  \\
\end{split}
\vspace{-8pt}
\end{IEEEeqnarray}
\text{subject to \hspace{-3pt}\{}
\vspace{-7pt}
\begin{IEEEeqnarray}{ccc} 
\label{7}
{g_{b^\textnormal{D}_{0},t}} &=& \sum_{b_n \in B_n(b^\textnormal{D}_{0})}{f^p_{(b^\textnormal{D}_{0},b_n),t}}: (\lambda_{b^\textnormal{D}_0,t}^\textnormal{D})\\[-4pt]
\label{8}
{q_{b^\textnormal{D}_{0}}} &=& \sum_{b_n \in B_n(b^\textnormal{D}_{0})}{f^\textnormal{q}_{(b^\textnormal{D}_{0},b_n),t}}: (\lambda_{b^\textnormal{D}_0,t}^\textnormal{Dq})
\end{IEEEeqnarray}
\vspace{-12pt}
\begin{IEEEeqnarray}{ccc}
\label{9}
\begin{split}
g_{b^\textnormal{D},t} + \sum_{b_m \in B_m(b^\textnormal{D})}{f^\textnormal{p}_{(b^\textnormal{D},b_m),t}} +\sum_{r \in R}{g^\textnormal{D}_{r,t}} = d^\textnormal{p}_{b^\textnormal{D},t}+ \\[-3pt]\sum_{b_n \in B_n(b^\textnormal{D})}{f^\textnormal{p}_{(b^\textnormal{D},b_n),t}} :(\lambda_{{b^\textnormal{D},t}}^D); \hspace{3mm} \forall b^\textnormal{D} \in B^\textnormal{D}
\end{split}
\end{IEEEeqnarray}
\vspace{-10PT}
\begin{IEEEeqnarray}{ccc}
\label{91}
\begin{split}
q_{b^\textnormal{D},t} + \sum_{b_m \in B_m(b^\textnormal{D})}{f^\textnormal{q}_{(b^\textnormal{D},b_m),t}} = d^\textnormal{q}_{b^\textnormal{D},t}+\hspace{-8pt}\sum_{b_n \in B_n(b^\textnormal{D})}{f^\textnormal{q}_{(b^\textnormal{D},b_n),t}} \\[-4pt]: (\lambda_{{b^\textnormal{D},t}}^\textnormal{Dq}); \hspace{5mm} \forall b^\textnormal{D} \in B^\textnormal{D} 
\end{split}
\end{IEEEeqnarray}
\vspace{-13PT}
\begin{IEEEeqnarray}{ccc} 
\label{10}
  G_{i^\textnormal{D}}^{{\textnormal{min}}} &\leq&  g_{i^\textnormal{D},t} \leq G_{i^\textnormal{D}}^{{\textnormal{max}}}:(\underline{\delta}_{i^\textnormal{D},t}, \overline{\delta}_{i^\textnormal{D},t}); \forall i^\textnormal{D} \in I^\textnormal{D}\\
\label{14}
  Q_{i^\textnormal{D}}^{{\textnormal{min}}} &\leq&  q_{i^\textnormal{D},t} \leq Q_{i^\textnormal{D}}^{{\textnormal{max}}}:(\underline{\theta}_{i^\textnormal{D},t}, \overline{\theta}_{i^\textnormal{D},t});\forall i^\textnormal{D} \in I^\textnormal{D}\\
 \label{res}
 0 &\leq& \sum_{r \in R}{g^\textnormal{D}_{r,t}} \leq g^{\textnormal{D}_o}_{r,t} :(\underline{\alpha}_{r,t}^\textnormal{D},\overline{\alpha}_{r,t}^\textnormal{D}); \forall r \in R \\[-5pt]
 \label{12} 
 U_{b^\textnormal{D}}^{{\textnormal{min}}} &\leq&  u_{b^\textnormal{D},t} \leq  U_{b^\textnormal{D}}^{{\textnormal{min}}}:(\underline{\mu}_{b^\textnormal{D},t}, \overline{\mu}_{b^\textnormal{D},t}); \forall b^\textnormal{D} \in B^\textnormal{D} \\
 \label{12.1}
 -F_{(b^\textnormal{T}_c,b_0^\textnormal{D})}^{\textnormal{max}}&\leq& f_{(b^\textnormal{T}_c,b_o^D),t} \leq F_{(b^\textnormal{T}_c,b_0^\textnormal{D})}^{\textnormal{max}} : (\underline{\tau}_{(b^\textnormal{T}_c,b_0^\textnormal{D})}, \overline{\tau}_{(b^\textnormal{T}_c,b_0^\textnormal{D})})
 \vspace{-5pt}
\end{IEEEeqnarray}
\begin{IEEEeqnarray}{ccc} 
\label{13}
\begin{split}
\vspace{-5pt}
[S_{(b^\textnormal{D},b_1)}; f^\textnormal{p}_{(b^\textnormal{D},b_1),t}; f^\textnormal{q}_{(b^\textnormal{D},b_1),t}] \in K\\[-2pt]: ([\eta_{(b^\textnormal{D},b_1)}^\textnormal{s}; \eta_{(b^\textnormal{D},b_1),t}^\textnormal{p};\eta_{(b^\textnormal{D},b_1),t}^\textnormal{q}]) \in K^*; \forall b^\textnormal{D}, b_1 \in B^\textnormal{D}
\end{split}
\vspace{-10pt}
\end{IEEEeqnarray}
\begin{IEEEeqnarray}{ccc} 
\label{121}
\begin{split}
\sum_{b_m \in B_m(b^\textnormal{D})}&\hspace{-10PT}2(X_{(b^\textnormal{D},b_m)}f^\textnormal{p}_{(b^\textnormal{D},b_m),t} + x_{(b^\textnormal{D},b_m)}f^\textnormal{q}_{(b^\textnormal{D},b_m),t})+ u_{b^\textnormal{D},t}  \\[-8pt]&= \hspace{-5pt}\sum_{b_m \in B_m}\hspace{-5pt}u_{b_{m,t}}\hspace{-3pt}: (\beta_{(b^\textnormal{D},b_m),t}); \forall b^\textnormal{D} \in B^\textnormal{D}\}, t \in T
\end{split}
\end{IEEEeqnarray}
\end{subequations}

\subsection{Formulation for SLMF Problem}
The SLMF problem described in Fig.~\ref{joint} is formulated as:
\begin{subequations}
\label{slmf_o}
\begin{IEEEeqnarray}{C}
\text{max} \hspace{10pt}\textnormal{Eq.}~\eqref{20}
\vspace{-10pt}
\end{IEEEeqnarray}
\vspace{-10pt}
\textnormal{subject to}
\begin{IEEEeqnarray}{rCl}
\textnormal{Eq.}~\eqref{21}
\vspace{-10pt}
\end{IEEEeqnarray}
\vspace{-10pt}
\begin{IEEEeqnarray}{C}
\lambda^\textnormal{T}_{b^\textnormal{T},t} \in \textnormal{arg \{Eq.}~\eqref{lmnopq}\}\\
\lambda^\textnormal{D}_{b^\textnormal{D},t} \in \textnormal{arg\{Eq.}~\eqref{distribution}\}
\end{IEEEeqnarray}
\end{subequations}

\subsection{KKT Conditions for the MPEC Reformulation}
 We reformulate the SLMF problem, described in eq.~\eqref{slmf_o}, into an MPEC using KKT conditions of the followers in Section~\ref{reform}. This  subsection provides the KKT conditions for the WM and DM.
\subsubsection{KKT conditions for wholesale market}
Consider the following stationary and complementarity conditions for \eqref{lmnopq}:
\begin{subequations}
\label{KKT_trans}
\begin{IEEEeqnarray}{ccc} 
\label{t1}
\{C_{i^\textnormal{T}}-\lambda_{b(i^\textnormal{T}),t}^\textnormal{T}-\underline{\gamma}_{i^\textnormal{T},t}+\overline{\gamma}_{i^\textnormal{T},t} = 0; \hspace{3mm}\forall i^\textnormal{T} \in I^\textnormal{T}\\
\label{t2}
C_{r}-\lambda_{b(r),t}^\textnormal{T} -\underline{\alpha}_{r,t}^\textnormal{T} +\overline{\alpha}_{r,t}^\textnormal{T} = 0;\hspace{3mm}\forall r \in R
\vspace{-10pt}
\end{IEEEeqnarray}
\begin{IEEEeqnarray}{ccc} 
\label{t3}
\begin{split}
-\sum_{b_m \in B_m(b^\textnormal{T})}{\lambda_{b_m, t}^\textnormal{T}}+\sum_{b_n \in B_n(b^\textnormal{T})}\lambda_{b_n,t}^\textnormal{T}-\xi_{(b^\textnormal{T},b_1),t}-\underline{\delta}_{(b^\textnormal{T},b_1),t}+\\\overline{\delta}_{(b^\textnormal{T},b_1),t} = 0; \hspace{5mm}\forall b^\textnormal{T}, b_1 \in B^\textnormal{T}\\
\end{split}\\
\label{t4}
-\hspace{-14pt}\sum_{b_m \in B_m(b^\textnormal{T})}\hspace{-5pt}{\frac{\xi_{(b^\textnormal{T},b_1),t}}{x_{(b^\textnormal{T},b_1)}}}+\hspace{-9pt}\sum_{b_n \in B_n(b^\textnormal{T})}\hspace{-5pt}{\frac{\xi_{(b_T,b_1),t}}{x_{(b^\textnormal{T},b_1)}}}=0; \forall b^\textnormal{T},b_1 \in B^\textnormal{T}
\vspace{-10pt}
\end{IEEEeqnarray}
\begin{IEEEeqnarray}{ccc} 
\label{t5}
0 \leq g_{i^\textnormal{T},t}-G_{{i^\textnormal{T}}}^\textnormal{min} \;\bot\; \underline{\gamma}_{i^\textnormal{T},t} \geq 0; \hspace{3mm} \forall i^\textnormal{T} \in I^\textnormal{T}\\
\label{t5.5}
0 \leq g_{r,t}^\textnormal{T} \;\bot\; \underline{\alpha}_{r,t}^\textnormal{T} \geq 0; \hspace{17mm} \forall r \in R
\vspace{-10pt}
\end{IEEEeqnarray}
\begin{IEEEeqnarray}{ccc} 
\label{t5.6}
0 \leq g_{r,t}^{\textnormal{T}_o}-{g_{r,t}^\textnormal{T}} \;\bot\; \overline{\alpha}_{r,t}^{\textnormal{T}} \geq 0;\hspace{10mm} \forall r \in R\\
\label{t6}
0 \leq G_{{i^\textnormal{T}, t}}^\textnormal{max}-g_{i^\textnormal{T}, t} \;\bot\; \overline{\gamma}_{i^\textnormal{T},t} \geq 0;\hspace{9mm} \forall i^\textnormal{T} \in I^\textnormal{T}
\vspace{-10pt}
\end{IEEEeqnarray}
\begin{IEEEeqnarray}{ccc} 
\label{t7}
0 \leq f_{(b^\textnormal{T},b1),t} + F_{{(b^\textnormal{T},b1)}}^\textnormal{min}\;\bot\; \underline{\delta}_{(b^\textnormal{T},b1),t} \geq 0;\hspace{1mm} \forall b^\textnormal{T},b_1 \in B^\textnormal{T}\\
0 \leq F_{{(b^\textnormal{T},b1)}}^\textnormal{max}-f_{(b^\textnormal{T},b1),t} \;\bot\; \overline{\delta}_{(b^\textnormal{T},b1),t} \geq 0;\hspace{1mm} \forall b^\textnormal{T},b_1 \in B^\textnormal{T}\\
\label{t8}
\text{Equality Constraints: Eqs.}~\eqref{5}, \eqref{3} \};\forall t\in T
\vspace{-5pt}
\end{IEEEeqnarray}
\end{subequations}
where $\bot$ denotes orthogonality between the variables, and $b(i^\textnormal{T})$ is the bus containing the generator $i^\textnormal{T} \in I^\textnormal{T}$. 

\subsubsection{KKT conditions for distribution market}
Stationary and complementarity conditions for \eqref{distribution} are computed as follows:
\vspace{-5pt}
\begin{subequations}
\label{KKT_dist}
\begin{IEEEeqnarray}{ccc} 
\label{d1}
\lambda_{b^\textnormal{D}_0,t}^\textnormal{D} -\underline{\tau}_{(b^\textnormal{T}_c,b_0^\textnormal{D}),t}+\overline{\tau}_{(b^\textnormal{T}_c,b_0^\textnormal{D}),t}=\lambda^\textnormal{T}_{b^\textnormal{T}_c,t}; \hspace{3mm} \forall t \in T\\
\label{d2}
\lambda_{b^\textnormal{D}_0,t}^\textnormal{Dq} =0;\hspace{3mm} \forall t \in T
\vspace{-10pt}
\end{IEEEeqnarray}
\begin{IEEEeqnarray}{ccc} 
\label{d3}
C_{i^\textnormal{D}}-\lambda_{{b(i^\textnormal{D}),t}}^{\textnormal{D}}-\underline{\delta}_{i^\textnormal{D},t}+\overline{\delta}_{i^\textnormal{D},t} = 0; \forall i^\textnormal{D} \in I^\textnormal{D}, t \in T\\
\label{d4}
C_{r} -\lambda_{{b(r),t}}^{\textnormal{D}}-\underline{\alpha}_{r,t}^\textnormal{D} +\overline{\alpha}_{r,t}^\textnormal{D} = 0; \forall r \in R; t \in T
\vspace{-10pt}
\end{IEEEeqnarray}
\begin{IEEEeqnarray}{ccc} 
\label{d5}
-\lambda_{{b(i^\textnormal{D}),t}}^{\textnormal{Dq}}-\underline{\theta}_{i^\textnormal{D},t}+\overline{\theta}_{i^\textnormal{D},t} = 0;\forall i^\textnormal{D} \in I^\textnormal{D}, t \in T\\
\label{d6}
\begin{split}
-\sum_{b_n \in B_n(b^\textnormal{D})}{\lambda_{{b_n,t}}^{\textnormal{D}}}+\lambda_{{b^\textnormal{D},t}}^{\textnormal{D}}-2\beta_{(b^\textnormal{D},b_m),t}X_{(b^\textnormal{D},b_m)}\\[-10pt]-\eta_{(b^\textnormal{D},b_m),t}^\textnormal{p} = 0;\forall b^\textnormal{D} \in B^\textnormal{D}, t \in T\\
\end{split}
\vspace{-10pt}
\end{IEEEeqnarray}
\begin{IEEEeqnarray}{ccc} 
\label{d7}
\begin{split}
-\sum_{b_n \in B_n(b^\textnormal{D})}\lambda_{{b_n,t}}^{\textnormal{Dq}}+\lambda_{{b^\textnormal{D},t}}^{\textnormal{Dq}}-2\beta_{(b^\textnormal{D},b_m),t}x_{(b^\textnormal{D},b_m)}\\[-10pt]-\eta_{(b^\textnormal{D},b_m),t}^\textnormal{q} = 0;\forall b^\textnormal{D} \in B^\textnormal{D}, t \in T\\
\end{split}\\
\label{d8}
\begin{split}
-\underline{\mu}_{b^\textnormal{D},t}+\overline{\mu}_{b^\textnormal{D},t}+ \sum_{b_m \in B_m(b^\textnormal{D})}\beta_{(b^\textnormal{D},b_m),t}\\-\beta_{(b^\textnormal{D},b_n),t}=0;\forall b^\textnormal{D} \in B^\textnormal{D}, t \in T\\
\end{split}
\vspace{-10pt}
\end{IEEEeqnarray}
\begin{IEEEeqnarray}{ccc} 
\label{d9}
0 \leq g_{i^\textnormal{D}, t}-G_{{i^\textnormal{D}, t}}^\textnormal{{min}} \;\bot\; \underline{\delta}_{i^\textnormal{D},t} \geq 0;\forall i^\textnormal{D} \in I^\textnormal{D}, t \in T\\
\label{d10}
0 \leq G_{{i^\textnormal{D}, t}}^\textnormal{{max}}-g_{i^\textnormal{D}, t} \;\bot\; \overline{\delta}_{i^\textnormal{D},t} \geq 0;\forall i^\textnormal{D} \in I^\textnormal{D}, t \in T
\vspace{-10pt}
\end{IEEEeqnarray}
\begin{IEEEeqnarray}{ccc} 
\label{d11}
0 \leq q_{i^\textnormal{D}, t}-Q_{{i^\textnormal{D}, t}}^\textnormal{{min}} \;\bot\; \underline{\theta}_{i^\textnormal{D},t} \geq 0;\forall i^\textnormal{D} \in I^\textnormal{D}, t \in T\\
\label{d12}
0 \leq Q_{{i^\textnormal{D}, t}}^\textnormal{{min}}-q_{i^\textnormal{D}, t} \;\bot\; \overline{\theta}_{i^\textnormal{D},t} \geq 0;\forall i^\textnormal{D} \in I^\textnormal{D}, t \in T
\vspace{-10pt}
\end{IEEEeqnarray}
\begin{IEEEeqnarray}{ccc} 
\label{d13}
0 \leq u_{b^\textnormal{D}, t}-U_{{b^\textnormal{D}, t}}^\textnormal{{min}} \;\bot\; \underline{\mu}_{b^\textnormal{D},t} \geq 0;\forall b^\textnormal{D} \in B^\textnormal{D}, t \in T\\
\label{d14}
0 \leq U_{{b^\textnormal{D}, t}}^\textnormal{{max}}-u_{b^\textnormal{D}, t} \;\bot\; \overline{\mu}_{b^\textnormal{D},t} \geq 0;\forall b^\textnormal{D} \in B^\textnormal{D}, t \in T
\vspace{-10pt}
\end{IEEEeqnarray}
\begin{IEEEeqnarray}{ccc} 
\label{d14.1}
0 \leq g_{b^\textnormal{D}_{0},t} \;\bot\;\underline{\tau}_{(b^\textnormal{T}_c,b_0^\textnormal{D}),t} \geq 0; \forall t \in T\\ 
\label{d14.2}
0 \leq F_{(b^\textnormal{T}_c,b_0^\textnormal{D})}^{\textnormal{max}} -g_{b^\textnormal{D}_{0},t} \;\bot\; \overline{\tau}_{(b^\textnormal{T}_c,b_0^\textnormal{D}),t} \geq 0; \forall t \in T
\vspace{-10pt}
\end{IEEEeqnarray}
\begin{IEEEeqnarray}{ccc} 
\label{d14.5}
0 \leq g_{r,t}^{\textnormal{D}_o}-\sum_{r \in R}{g_{r,t}^\textnormal{D}} \;\bot\; \overline{\alpha}_{r,t}^{\textnormal{D}} \geq 0; \forall r \in R, t \in T\\[-3pt]
\vspace{-10pt}
\label{d14.6}
0 \leq \sum_{r \in R}{g_{r,t}^\textnormal{D}} \;\bot\; \underline{\alpha}_{r,t}^{\textnormal{D}} \geq 0;\forall r \in R, t \in T
\end{IEEEeqnarray}
\begin{IEEEeqnarray}{ccc} 
\begin{split}
[S_{(b^\textnormal{D},b_1)}; f^\textnormal{p}_{(b^\textnormal{D},b_1),t}; f^\textnormal{q}_{(b^\textnormal{D},b_1),t}] \;\bot\;[\eta_{(b^\textnormal{D},b_1)}^\textnormal{s}; \eta_{(b^\textnormal{D},b_1),t}^\textnormal{p};\\\eta_{(b^\textnormal{D},b_1),t}^\textnormal{q}]; \forall b^\textnormal{D}, b_1 \in B^\textnormal{D}, t \in T
\end{split}\\
\label{d15}
\text{Equality Constraints: Eqs.}~\eqref{7} - \eqref{91}, \eqref{121}
\end{IEEEeqnarray}
\end{subequations}

\subsection{MPEC Reformulation }
\label{MPEC_reform_1}
In this section, we incorporate the upper-level (UL) problem defined in eq.~\eqref{leader}, and the KKT conditions for the two lower level (LL) problems, eqs.~\eqref{KKT_trans} and \eqref{KKT_dist}, to formulate single-level MPECs for joint and sequential cases, as follows: 
\label{reform}
\subsubsection{Joint Case}
\label{joint_reform}
We formulate the MPEC for the joint SLMF game described in Fig.~\ref{joint}, as follows:
\begin{subequations}
\label{seq_form}
\begin{IEEEeqnarray}{l}
\label{jm}
\text{max} \; \textnormal{Eq.}~\eqref{20}
\vspace{-10pt}
\end{IEEEeqnarray}
\vspace{-10pt}
subject to
\begin{IEEEeqnarray}{lll}
\text{UL Constraint:} &\hspace{5pt} \textnormal{Eq.} & \eqref{21}
\\
\text{LL KKT Conditions:} &\hspace{5pt} \textnormal{Eqs.} & \eqref{t1} - \eqref{d15}
\end{IEEEeqnarray}
\end{subequations}

\subsubsection{Sequential Case}
Similarly, the MPEC reformulation for the sequential SLMF game described in Fig.~\ref{seq}, is given as:
\vspace{-5pt}
\begin{subequations}
\begin{IEEEeqnarray}{C}
\label{20_100}
\begin{split}
\text{max} \sum_{r \in R, t \in T}{(\lambda^\textnormal{T}_{b^\textnormal{T},t} } - C_{r}) g^{\textnormal{T}_o}_{r,t}+ \hspace{-10pt} \sum_{r \in R, t \in T}{(\wp_{b^\textnormal{D},t}} - C_{r}) g^{\textnormal{D}_o}_{r,t}
\end{split}
\end{IEEEeqnarray}
\vspace{-10 pt}
subject to
\vspace{-5pt}
\begin{IEEEeqnarray}{lll}
\label{21_1}
\textnormal{UL Constraint:} &\hspace{5pt} \textnormal{Eq}. &\eqref{21}\\
\label{21_3}
\textnormal{LL KKT Conditions:}&\hspace{5pt} \textnormal{Eqs.} &\eqref{t1} - \eqref{t8}
\end{IEEEeqnarray}
\end{subequations}
where the random variable $\wp_{b^\textnormal{D},t}$ follows a Gaussian distribution, i.e. $\wp_{b^\textnormal{D},t} \sim \mathcal{N}(\overline{\lambda}^\textnormal{D}_{b^\textnormal{D},t},\,\sigma^2_{b^\textnormal{D},t})$, with a mean DLMP at bus $b^\textnormal{D}$ during time interval $t$ given by $\overline{\lambda}^\textnormal{D}_{b^\textnormal{D},t}$ and the  standard deviation given by $\sigma_{b^\textnormal{D},t}$. The CC reformulation of the term with $\wp_{b^\textnormal{D},t}$ in eq.~\eqref{20_100} is then given by, \cite{boyd}: 
\vspace{-5 pt}
\begin{IEEEeqnarray}{C}
\textnormal{Let} \hspace{10 pt}(\wp_{b^\textnormal{D},t})g^{\textnormal{D}_o}_{r,t} \geq z_{r,t} \hspace{5 pt}; \hspace{5 pt} \forall \hspace{5 pt} r \in R, t \in T  \\
\implies \mathbb{P}[\wp_{b^\textnormal{D},t} \geq \frac{z_{r,t}}{g^{\textnormal{D}_o}_{r,t}}] \geq \ 1- \epsilon \vspace{-7 pt}\\
\label{cc}
\implies g^{\textnormal{D}_o}_{r,t}(\overline{\lambda}^\textnormal{D}_{b^\textnormal{D},t} -\phi^{-1}(1 - \epsilon)\sigma_{b^\textnormal{D},t}) \geq z_{r,t}
\end{IEEEeqnarray}
where $z_{r,t}$ is an auxiliary variable, $\epsilon$ is the confidence interval for the deterministic value of random variable, and $\phi^{-1}(.)$ is the quantile function of the standard normal distribution.
Hence, the MPEC reformulation for the first phase (i.e. interactions with the WM in Fig.~\ref{seq}) of the sequential SLMF game can be written as:
\vspace{-5pt}
\begin{subequations}
\label{cc_opt}
\begin{IEEEeqnarray}{C}
\text{max}  \sum_{r \in R, t \in T}{\hspace{-10pt}(\lambda^\textnormal{T}_{b^\textnormal{T},t} } - C_{r}) g^{\textnormal{T}_o}_{r,t}+\hspace{-10pt} \sum_{r \in R, t \in T}\hspace{-8pt}[({-C_{r})g^{\textnormal{D}_o}_{r,t}} + z_{r,t}]
\end{IEEEeqnarray}
\vspace{-15 pt}
subject to
\vspace{-5 pt}
\begin{IEEEeqnarray}{lll}
\label{cc_ul}
& \textnormal{Eqs}. &\hspace{5pt}\eqref{21_1}, \eqref{21_3}, \eqref{cc} 
\end{IEEEeqnarray}
\end{subequations}

We observe from eq.~\eqref{cc} that the CC formulation is identical to the formulation of $\textnormal{VaR}_\epsilon$ for any random variable. For a generic random variable \textbf{Y}, we define:
\begin{subequations}
\begin{IEEEeqnarray}{C}
\label{Var}
\textnormal{VaR}_{\epsilon}(\bold{Y}) := \text{min} \{y | \mathbb{P}(\bold{Y} \leq y) \geq \epsilon\} \\
\label{Var_cc}
\mathbb{P}(\bold{Y} \leq y) \geq \epsilon \xleftrightarrow{} \mu + \phi^{-1}(\epsilon)\sigma \leq y
\end{IEEEeqnarray}
\end{subequations}

Due to the robustness and coherence of CVaR, explained in Section~\ref{intro}, we employ it as follows:
\begin{IEEEeqnarray}{C}
\label{CVar}
\textnormal{CVaR}_{\epsilon}(\bold{Y}) := \mathbb{E} \{\bold{Y} | \bold{Y} \geq \textnormal{VaR}_{\epsilon}(\bold{Y})\}
\end{IEEEeqnarray}
Hence, for an independent random variable $\wp_{b^\textnormal{D},t}$, eq.~\eqref{cc} can be reformulated in terms of $\textnormal{CVaR}_{\epsilon}$ as follows:
\begin{IEEEeqnarray}{C}
\textnormal{Let} \hspace{10 pt}(\wp_{b^\textnormal{D},t})g^{\textnormal{D}_o}_{r,t} \geq m_{r,t} \hspace{5 pt}; \hspace{5 pt} \forall \hspace{5 pt} r \in R, t \in T  \\
\implies \mathbb{E}[\wp_{b^\textnormal{D},t} | \wp_{b^\textnormal{D},t} \geq \frac{m_{r,t}}{g^{\textnormal{D}_o}_{r,t}}] \geq \ \epsilon\\
\implies \textnormal{CVaR}_{\epsilon}(\wp_{b^\textnormal{D},t}) \geq \frac{m_{r,t}}{g^{\textnormal{D}_o}_{r,t}}  \\
\label{cvar}
\implies g^{\textnormal{D}_o}_{r,t}(\overline{\lambda}^\textnormal{D}_{b^\textnormal{D},t} -\frac{\Phi(\phi^{-1}( \epsilon))}{1-\epsilon}\sigma_{b^\textnormal{D},t}) \geq m_{r,t}
\end{IEEEeqnarray}
where $m_{r,t}$ is an auxiliary variable, $\phi^{-1}(.)$ is an inverse cumulative distribution function of a standard Gaussian, and $\Phi(.)$ is an associated probability density function. 
Hence, the MPEC reformulation for the first phase of sequential SLMF game can be written as:
\begin{subequations}
\label{seq_cvar}
\begin{IEEEeqnarray}{C}
\label{20_1}
\begin{split}
\text{max}\hspace{-5pt} \sum_{r \in R, t \in T}{\hspace{-10pt}(\lambda^\textnormal{T}_{b^\textnormal{T},t} } - C_{r}) g^{\textnormal{T}_o}_{r,t} +\hspace{-10pt} \sum_{r \in R, t \in T}\hspace{-8pt}[({-C_{r})g^{\textnormal{D}_o}_{r,t}} + m_{r,t}]
\end{split}
\end{IEEEeqnarray}
\vspace{-10pt}
subject to
\vspace{-5pt}
\begin{IEEEeqnarray}{lll}
& \textnormal{Eqs}. &\hspace{5pt} \eqref{21_1}, \eqref{21_3}, \eqref{cvar}
\end{IEEEeqnarray}
\end{subequations}

Let $g^{\textnormal{T*}}_{r,t}$ and $\lambda^{\textnormal{T*}}_{b^\textnormal{T},t}$ denote the market clearing outcome of the SLSF game between the strategic producer and WM, defined in eqs.~\eqref{cc_opt} and \eqref{seq_cvar}. As per the designed sequential market clearing mechanism, these outcomes would be communicated to the producer and DM, which then parametrize their SLSF game in this outcome as follows:
\begin{subequations}
\label{PSLSF}
\begin{IEEEeqnarray}{rCl}
\label{last}
\text{max}\sum_{r \in R, t \in T}{(\lambda^\textnormal{D}_{b^\textnormal{D},t}} - C_{r}) g^{\textnormal{D}_o}_{r,t}
\end{IEEEeqnarray}
\vspace{-8 pt}
subject to
\begin{IEEEeqnarray}{lll}
\label{222222}
\textnormal{UL Constraint:}&\hspace{5 pt} G_r^{\textnormal{min}}  \leq & g^{\textnormal{T*}}_{r,t} + g^{\textnormal{D}_o}_{r,t} \leq G_r^{\textnormal{max}}\\
\textnormal{LL KKT Conditions:}& \hspace{5pt}{\textnormal{Eqs}}.& \eqref{d2} - \eqref{d15}
\end{IEEEeqnarray}
\end{subequations}

\textbf{Remark 1}. For the joint and sequential cases with the regulated distribution environment, eqs.~\eqref{jm}, \eqref{20_100}, and \eqref{last} are modified to replace uncertain DLMPs given by $\lambda^\textnormal{D}_{b^\textnormal{D},t}$ and $\wp_{b^\textnormal{D},t}$ with tariff $\pi^\textnormal{D}_t$.

\section{Solution Technique}
\label{sol_tech}
\begin{figure}[!t]
\centering
\includegraphics[width=3.3in]{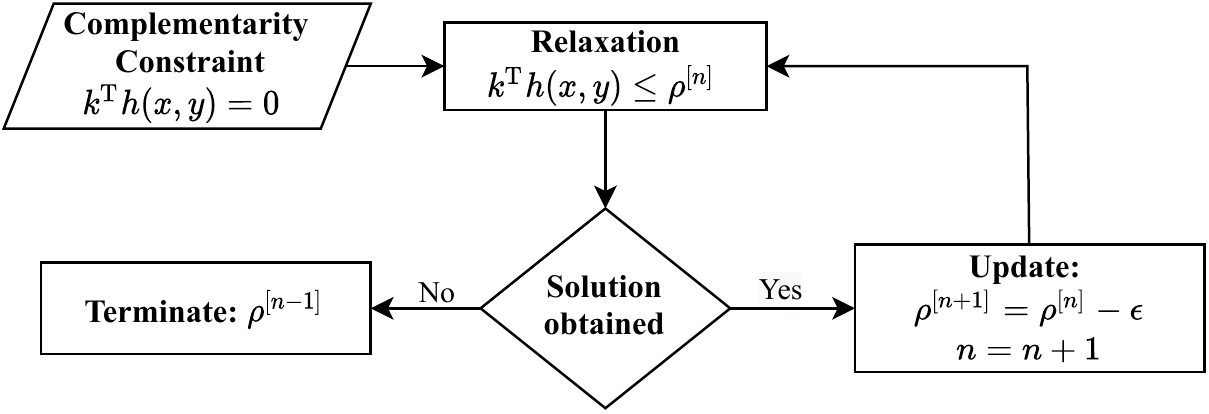}
\vspace{-10pt}
\caption{\small{An illustrative diagram for Scholte's relaxation scheme.}}
\label{scholtes}
\end{figure}

To elaborate the global relaxation scheme by Scholtes, we define a generic MPEC of the form:
\begin{subequations} \label{eq:gen_mpec}
\begin{IEEEeqnarray}{c}
\text{max} \; f(x,y)
\end{IEEEeqnarray}
subject to
\vspace{-20pt}
\begin{IEEEeqnarray}{c}
g(x,y) \geq 0\\
h(x,y) \geq 0 \; \text{and} \;k \geq 0\\
k^\textnormal{T} h(x,y) = 0 \label{comp},
\end{IEEEeqnarray}
\end{subequations}
where \eqref{eq:gen_mpec} can accommodate SLMF formulations in eqs.~\eqref{seq_form}, \eqref{cc_opt} and \eqref{seq_cvar}, and \eqref{cc_opt} and \eqref{PSLSF}.  This methodology relaxes the complementarity constraint, eq. (\ref{comp}), using a small user-defined relaxation value $\rho$, such that $k^\textnormal{T} h(x,y) \leq \rho$. The resulting NLP formulation is iteratively solved such that in each iteration, the value of $\rho$ is decreased from its previous value till $\rho \rightarrow 0$. The associated solution of the relaxed NLP where $\rho \approx 0$ is the solution of MPEC. An illustration for this iterative scheme is shown in Fig.~\ref{scholtes}.

\begin{center}
\begin{table}[t]
\caption{Market clearance models for T\&D markets.}
\centering
\begin{tabular}{ c | c | c | c }
\hline
\hline
  & GCT & Dist. Environment & Risk Measure  \\
\hline 
Case A & Simultaneous & Deregulated & N/A   \\ 
Case B & N/A & Regulated & N/A  \\
Case C & Sequential &Deregulated & VaR \\
Case D & Sequential &Deregulated & CVaR \\
\hline
\end{tabular}
\label{cases_table}
\vspace{-15pt}
\end{table}
\end{center}

\section{Case study}
\begin{figure}[!t]
\vspace{-10pt}
\centering
\includegraphics[height=2.5in,width=3.8in]{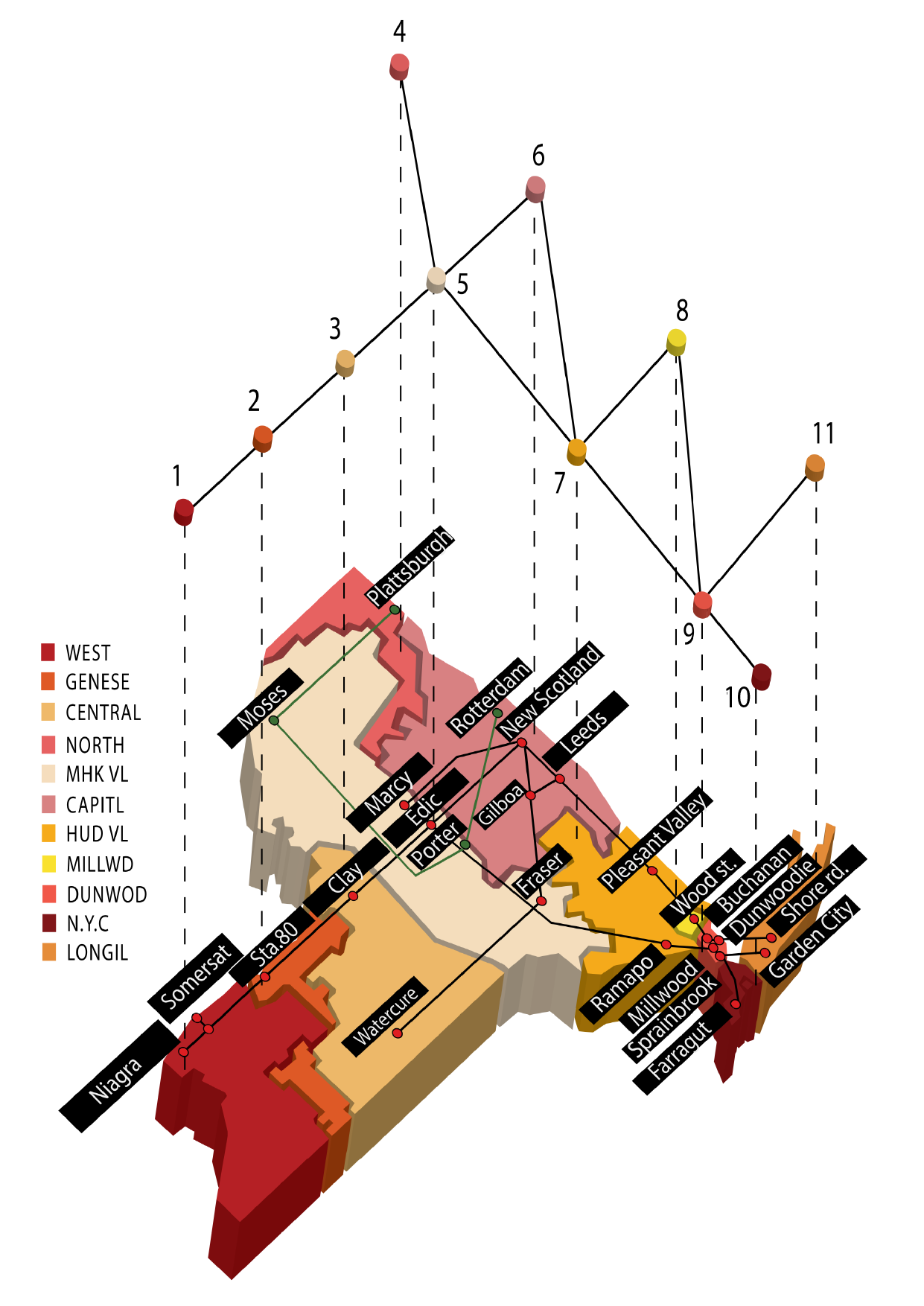}
\vspace{-25pt}
\caption{\small{A diagram of the 11-zone NYISO transmission system  with the distribution system  connected to buses 10 and 11.}}
\label{NYISO}
\end{figure}

\begin{figure}[!t]
\vspace{-15pt}
\centering
\includegraphics[height=2.5in, width=3.8in]{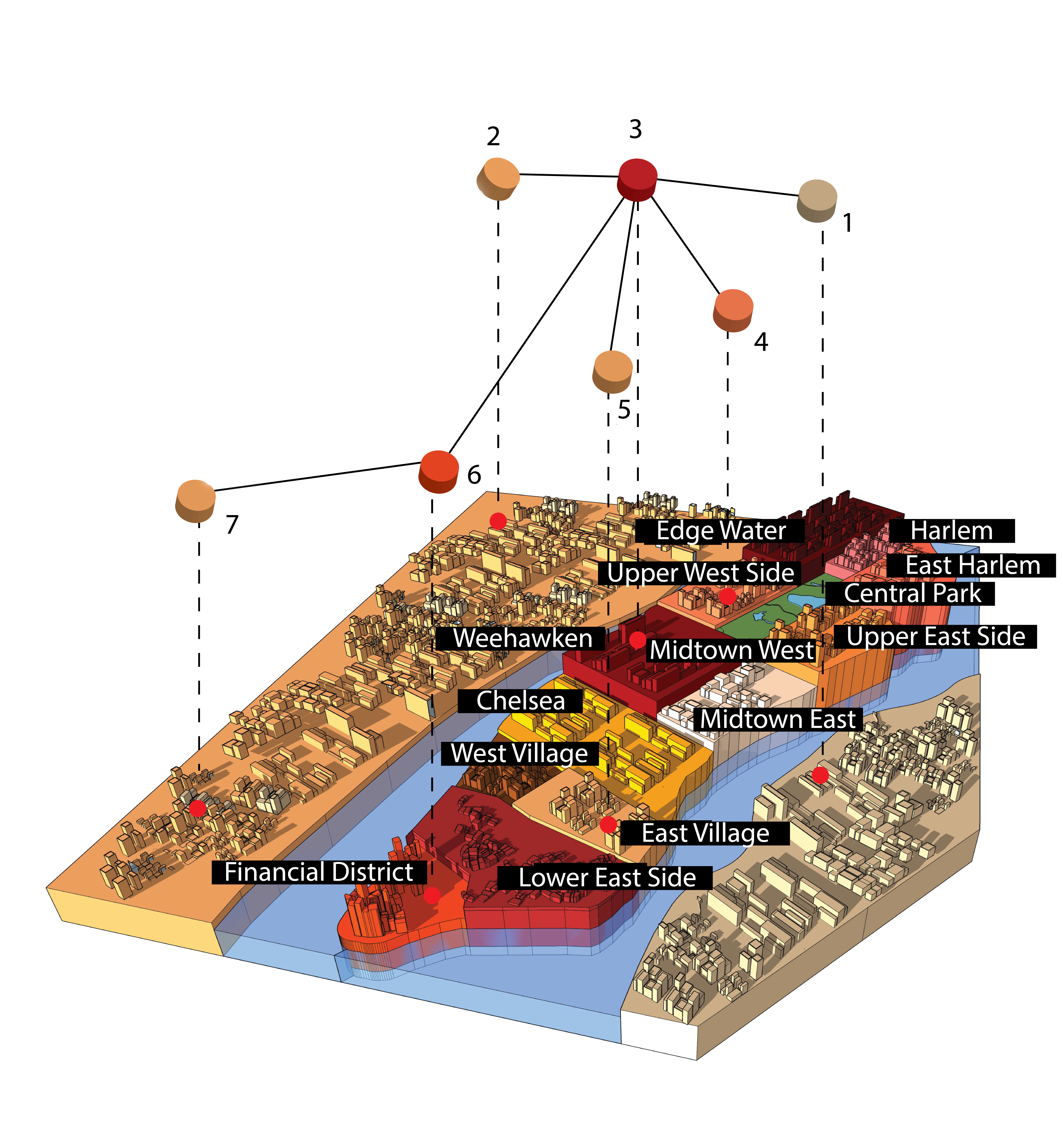}
\vspace{-30pt}
\caption{\small{A diagram of the 7-bus Manhattan distribution system with the transmission system connected to buses 1 and 6.}}
\label{Manhattan}
\vspace{-20pt}
\end{figure}

The case study uses  the 11-zone New York ISO (NYISO) transmission system \cite{goldbook} for the WM and the 7-bus Manhattan distribution system \cite{samrat} for the DM. Figs.~\ref{NYISO} and \ref{Manhattan} display these two systems. The total generation portfolio of NYISO is  39.3 GW (12.6 GW for oil, 12.8 GW for gas, 5.7 GW for hydro, 5.4 GW for nuclear, 1.7 GW for wind, 0.84 GW for coal, 0.03 GW for solar, and 0.22 GW for waste generators) and  the Manhattan system has one  strategically acting conventional gas generator of 716 MW. We model three interconnecting lines between the transmission and  distribution networks (i.e. NYC (bus \# 10) to bus \# 1, Long Island (bus \# 11) to bus \# 1, and Long Island (bus \# 11) to bus \# 6). Simulations are carried for four seasons using representative days for each season with the data available in \cite{goldbook}. 

Using this data and the formulations  in Section~\ref{MPEC_reform_1}, we define four cases for this case study, summarized in Table~\ref{cases_table}. Case A is the joint market clearance with the competitive DM, formulated in eq.~\eqref{seq_form}. Case B is the joint market clearance with a regulated distribution environment as in Remark 1. Case C is the sequential market clearance with CC, given by eq.~\eqref{cc_opt}, and Case D is the sequential market clearance with CVaR, see eq.~\eqref{seq_cvar}. For the case of the regulated distribution environment, we use time-of-use-tariffs for ConEdison available in \cite{tariff}. The values of these tariffs are set based on peak load hours, i.e. for peak hours (8 a.m to midnight) tariff is 21.97 c/kWh, whereas for off-peak hours (midnight to 8 a.m) tariff is 1.55 c/kWh. The value of constraint violations ($\epsilon$) is set to 5\% in all cases. 

\subsection{Joint Case}
\begin{figure}[!t]
\begin{center}
\includegraphics[height=2.3in, width=3.8in]{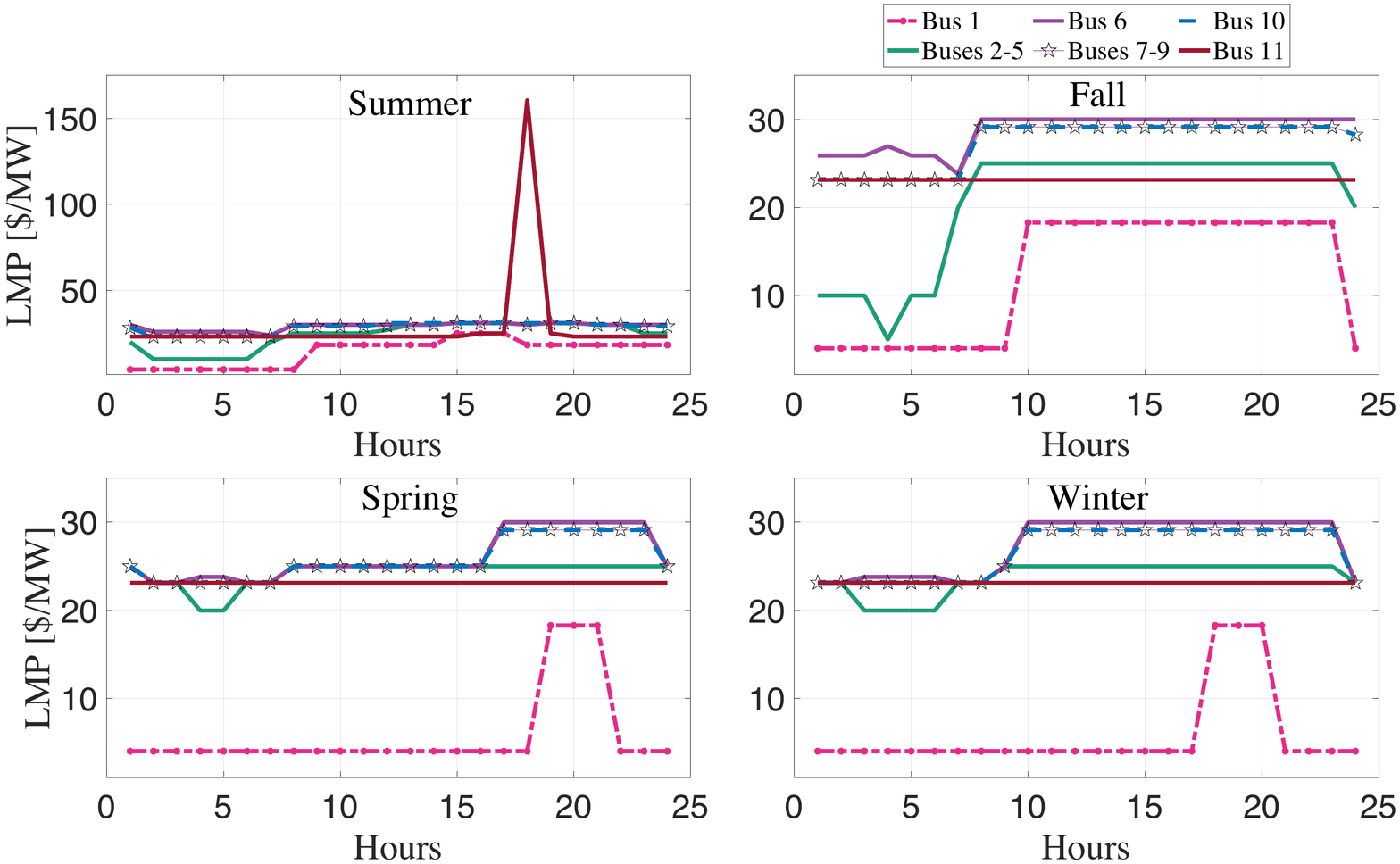}
\caption{\small{LMPs of NYISO for different seasons (Case A). }}
\label{jc_d_lmp}
\vspace{-20pt}
\end{center}
\end{figure}

\begin{figure}[!t]
\begin{center}
\includegraphics[height=2.3in, width=3.8in]{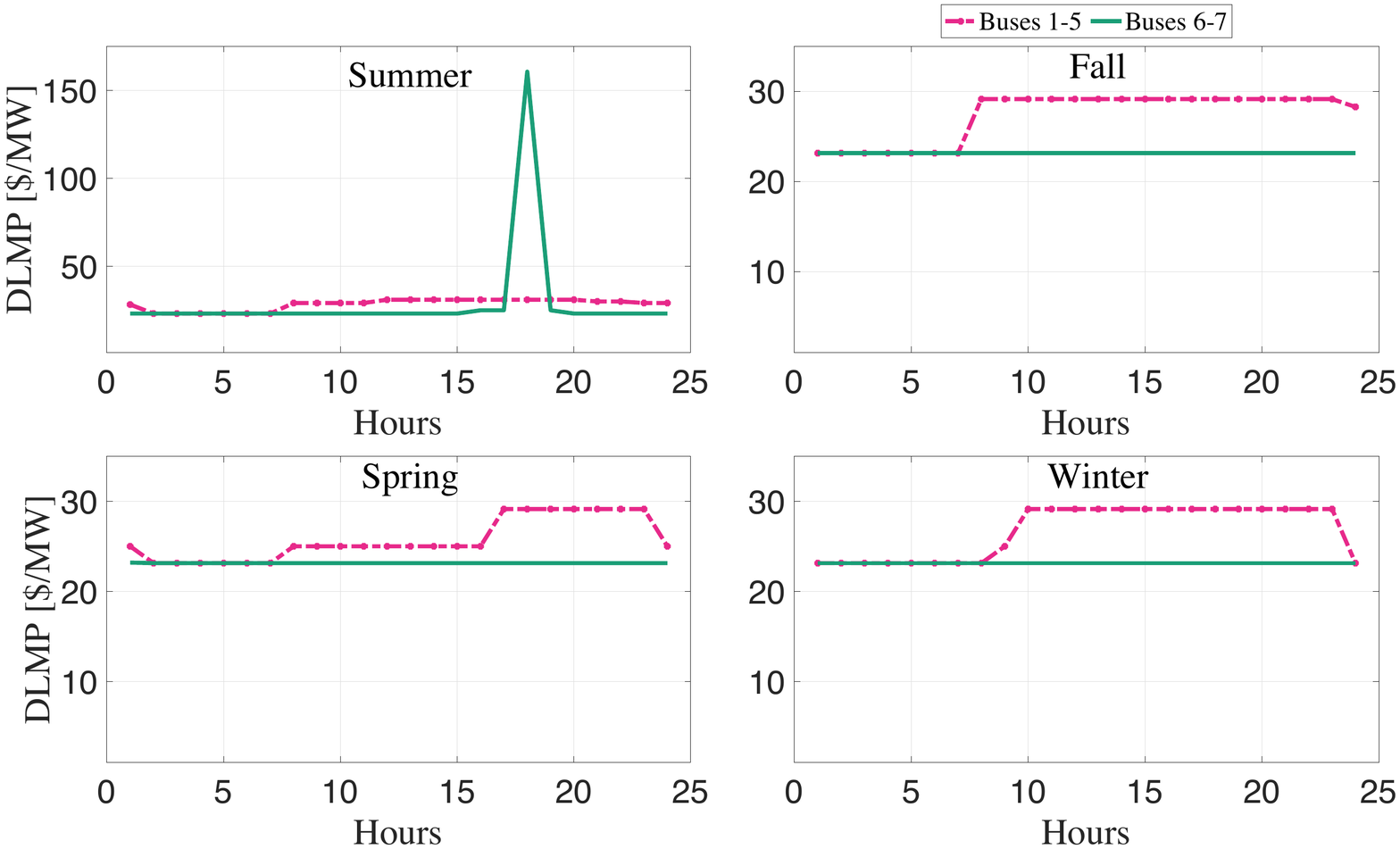}
\caption{\small{DLMPs of Manhattan for different seasons (Case A). }}
\label{jc_d_dlmp}
\vspace{-20pt}
\end{center}
\end{figure}

Figs.~\ref{jc_d_lmp} and \ref{jc_d_dlmp} compare
daily LMPs and DLMPs for four seasons, while modeling a competitive DM for Case A. There are 6 distinct LMP and 2 distinct DLMP profiles that are consistently observed across different seasons and vary substantially throughout each representative day. A high penetration of oil and gas generators in the large load zones of NYC (bus \# 10) and Long Island (bus \# 11) also drive the LMP patterns in the transmission nodes between Hudson Valley (bus \# 7) and Long Island (bus \# 11). 

\begin{figure}[!t]
\begin{center}
\includegraphics[height=2.3in, width=3.8in]{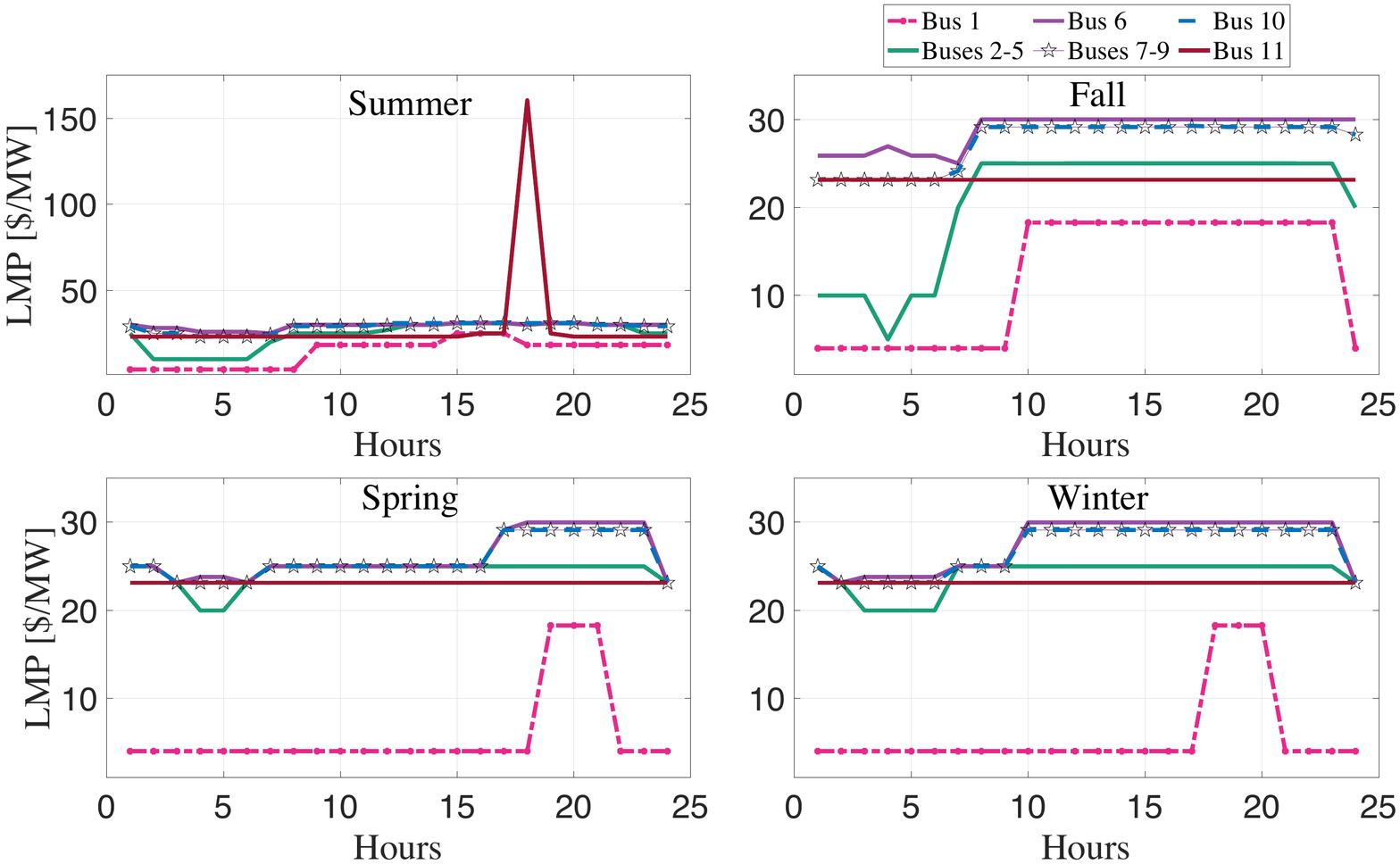}
\caption{\small{LMPs of NYISO for different seasons (Case B). }}
\label{jc_r_lmp}
\vspace{-20pt}
\end{center}
\end{figure}

\begin{figure}[!t]
\begin{center}
\includegraphics[height=2.8in, width=3.8in]{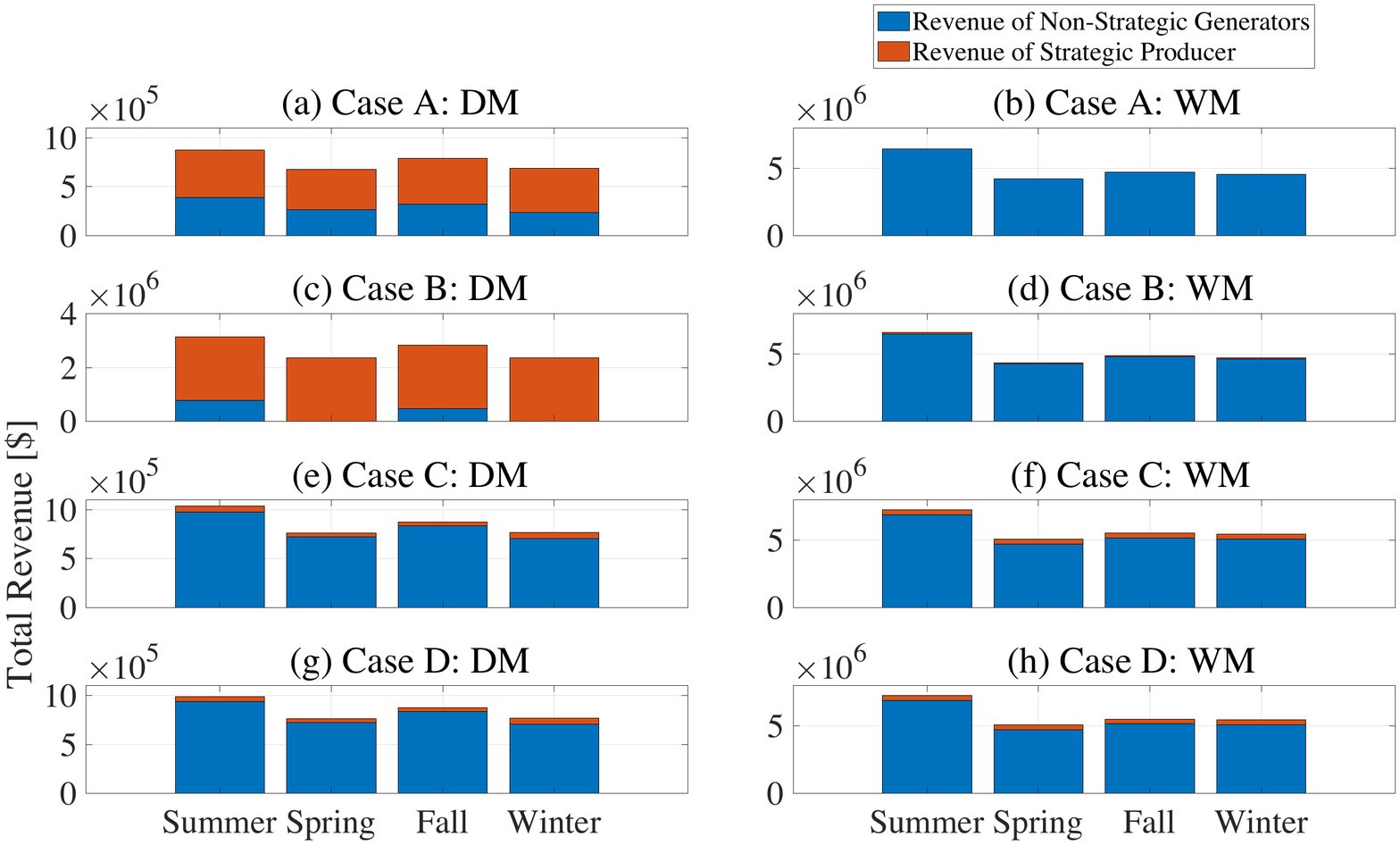}
\vspace{-20pt}
\caption{\small{Revenue of  market participants of T\&D markets. }}
\label{rev}
\vspace{-20pt}
\end{center}
\end{figure}

For comparison, Fig. \ref{jc_r_lmp} reports NYISO LMPs for Case B. While the general trend for LMPs is similar to Case A in Fig.~\ref{jc_d_lmp},  the patterns for LMPs in NYC (bus \# 10) are different in these two cases due to the deviations between the tariff and DLMP at bus \# 5 in the Manhattan system. Since the strategic producer optimizes its offer to the two markets based on these LMPs and DLMPs,   transitioning  from the current regulated environment  to a competitive DM will affect the ratio between the  power provision at distribution and transmission levels. Accordingly, Fig.~\ref{rev} compares revenues collected by strategic and   non-strategic producers from the services provided to both transmission and distribution networks. Thus, comparing Figs.~\ref{rev} (a) with (b), and (c) with (d), the strategic producer collects most of its revenue from the DM and its participation in the WM is minimal in Cases A and B. The regulated  tariff forces the strategic producer to increase its participation at the distribution level due to a very high time-of-use tariff in Manhattan, as seen in Figs.~\ref{rev} (a) and (c).

\begin{figure}[!t]
\begin{center}
\includegraphics[height=2.3in, width=3.8in]{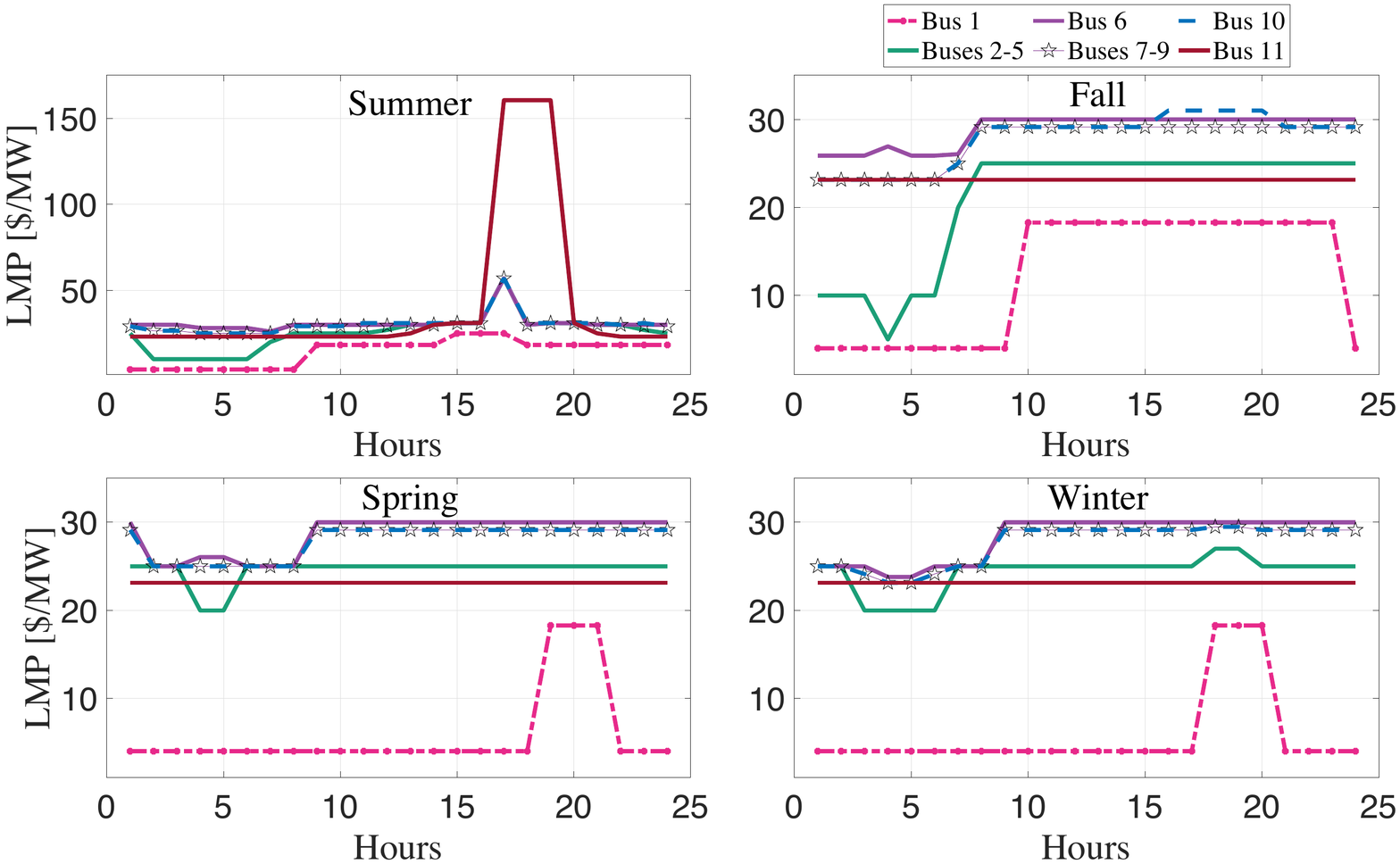}
\caption{\small{LMPs of NYISO for different seasons (Case C). }}
\label{sc_d_lmp}
\vspace{-20pt}
\end{center}
\end{figure}

\begin{figure}[!t]
\begin{center}
\includegraphics[height=2.3in, width=3.8in]{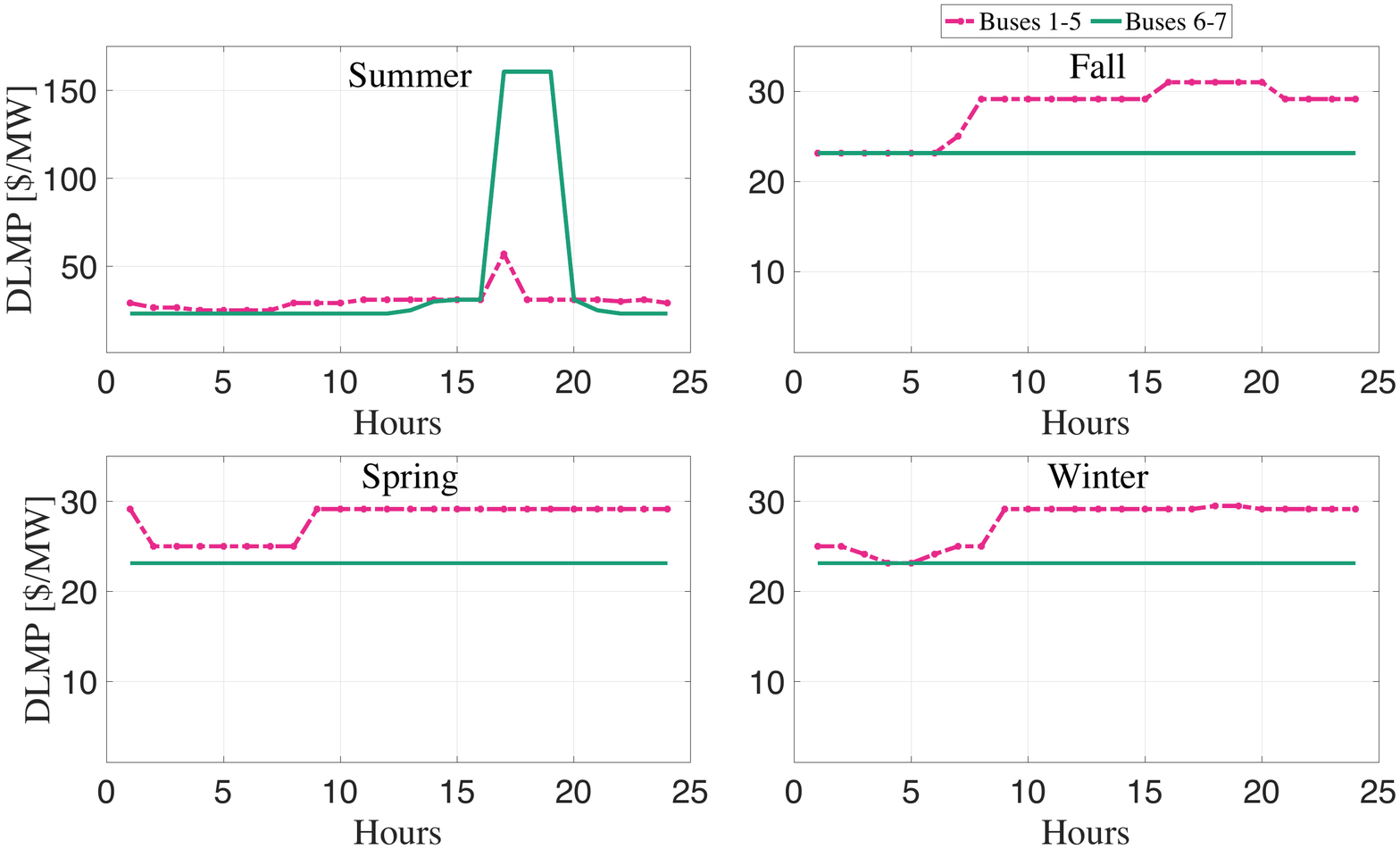}
\caption{\small{DLMPs of Manhattan for different seasons (Case C). }}
\label{sc_d_dlmp}
\vspace{-20pt}
\end{center}
\end{figure}

\begin{figure}[!t]
\begin{center}
\includegraphics[height=2.3in, width=3.8in]{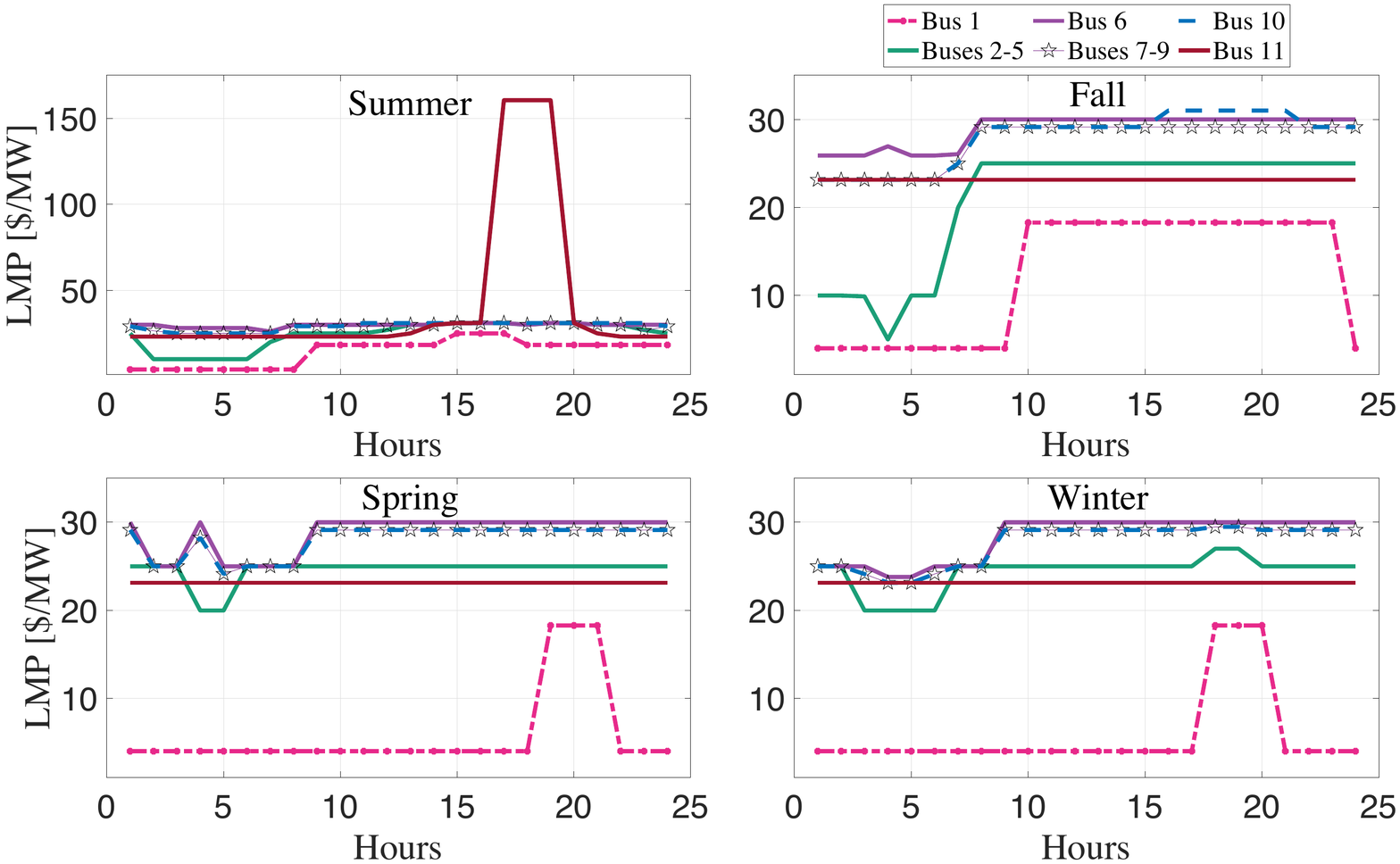}
\caption{\small{LMPs of NYISO for different seasons (Case D). }}
\label{sc_d_lmp_cvar}
\vspace{-18pt}
\end{center}
\end{figure}

\begin{figure}[!t]
\begin{center}
\includegraphics[height=2.3in, width=3.8in]{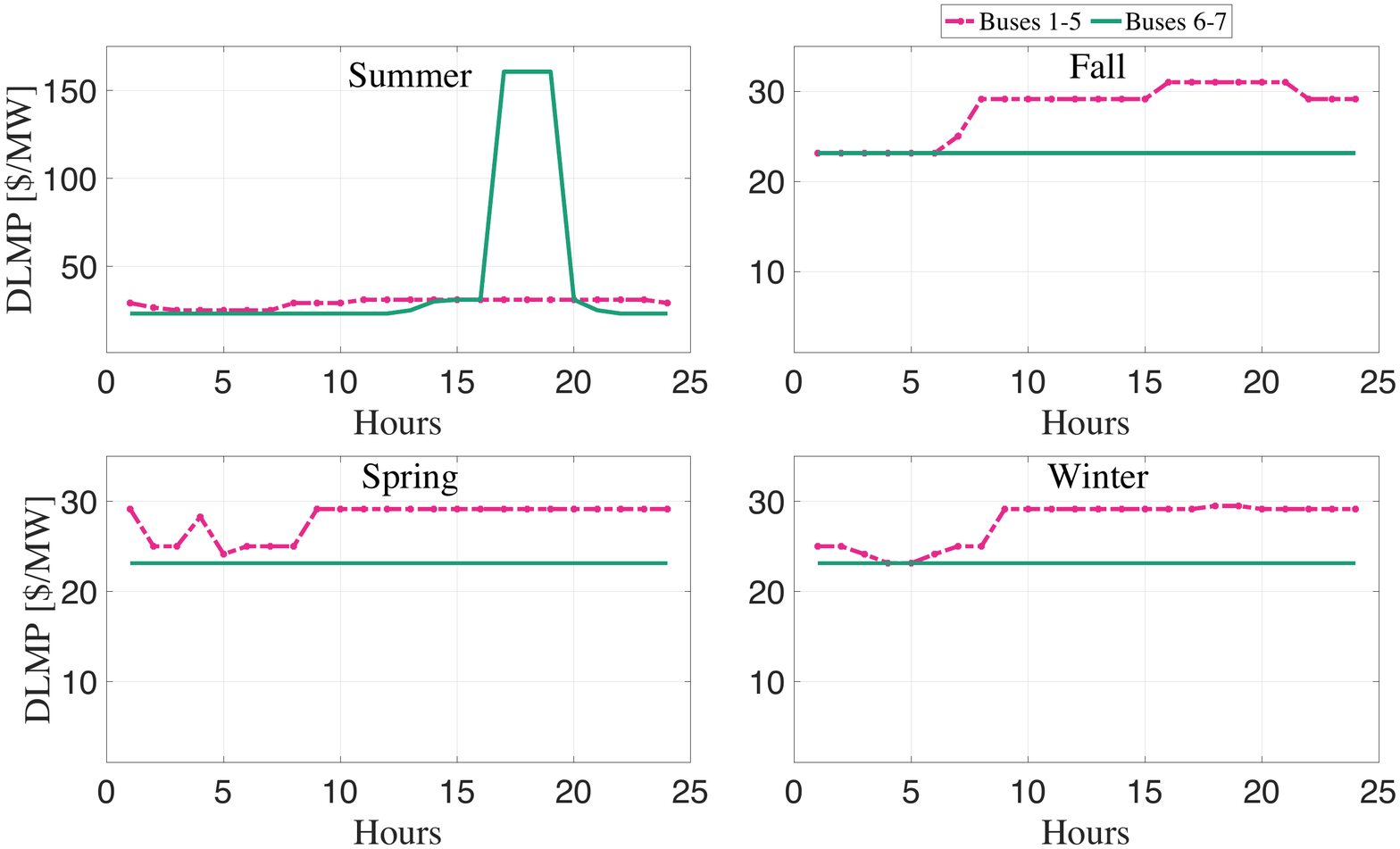}
\caption{\small{DLMPs of Manhattan for different seasons (Case D). }}
\label{sc_d_dlmp_cvar}
\vspace{-18pt}
\end{center}
\end{figure}

\subsection{Sequential Case}
The sequential market clearing model introduces a gap between the WM and DM GCTs, which makes it possible to arbitrage between the two markets. In the following simulations, we model this uncertainty using CCs, where $\sigma \approx 5.65\;\$$/MWh, computed from the historical  LMPs from NYISO. Hence, depending on the realization of random variable  $\wp_{b^D,t}$, the strategic producer chooses its generation offer for the first SLSF game as modeled in eqs.~\eqref{cc_opt} and \eqref{seq_cvar}, whereas the remaining capacity is reserved for the second parametrized SLSF game as modeled in eq.~\eqref{PSLSF}. 

Comparing Figs. \ref{sc_d_lmp} and \ref{sc_d_dlmp} (Case C) to Figs.~\ref{jc_d_lmp} and \ref{jc_d_dlmp} (Case A)  suggests that this arbitrage between the WM and DM leads to LMP variations in NYC (bus \#10) and within its proximity. As a result of this uncertainty, the strategic producer offers less than its full capacity in the DM and its revenue from the WM increases for all seasons, see Fig. \ref{rev} (e) and (f). Comparing Figs.~\ref{rev} (b) and (f), we observe a 98.4\% increase in the revenue of the producer from the WM for Case C. On the other hand, under the regulated tariff,  the results are the same as in the joint case, since the tariff is a priori known and does not provide opportunities for arbitrage or impose any risks. 

Given the uncertainty faced in Cases C and D, using CVaR makes the strategic producer more conservative in its market participation, which leads to lower revenues, see  Figs. \ref{rev} (g) and (h), than when uncertainty is treated by means of CC, as in Figs. \ref{rev} (e) and  (f). Similarly, the LMP and DLMP patterns for Case D are shown in Figs. \ref{sc_d_lmp_cvar} and \ref{sc_d_dlmp_cvar}.

We observe that the strategic producer generates maximum revenue in Case B due to a very high time-of-use tariff in Manhattan during the peak times. Case A forms the penultimate case in terms of revenue, whereas the least revenue is obtained for sequential case (specifically, Case D). We note that Case A reduces the total revenue of the strategic producer by 81.6\% as compared to Case B, whereas this reduction is 83.8\% while comparing Case A and Case D. 

\section{Conclusion}
This paper investigates risk-informed participation strategies for strategic producers in T\&D markets. This risk, emanating from the roll-out of distribution markets, is quantified by formulating and solving a risk-informed profit maximization problem in joint- and sequentially-cleared T\&D markets with non-identical GCTs. The formulations incorporate CCs and CVaR to internalize price uncertainties and exploit relaxation schemes for MPECs to solve the resulting SLMF game. To be consistent with the current market practices, this paper incorporated a regulated distribution environment, which is used for benchmark comparison with T\&D cases.  Relative to the regulated distribution environment, introducing a competitive DM tends to reduce profit opportunities for strategic producers due to the imposed uncertainty and risk. For instance, we observe that the uncertainty in DLMPs in the deregulated distribution environment reduces the profit at least by 81.6\% (compare Cases A \& B), and at most by 83.8\% (compare Cases A \& D). 

\bibliographystyle{IEEEtran}
\bibliography{references}{}

\end{document}